# SICRET: Supernova Ia Cosmology with truncated marginal neural Ratio EsTimation

Konstantin Karchev ,[1]★ Roberto Trotta[1,2] and Christoph Weniger[3]

[1]*Theoretical and Scientific Data Science Group, Scuola Internazionale Superiore di Studi Avanzati (SISSA), via Bonomea 265, I-34136 Trieste, Italy*
[2]*Astrophysics Group, Department of Physics, Blackett Lab, Imperial College London, Prince Consort Road, London SW7 2AZ, UK*
[3]*Gravitation Astroparticle Physics Amsterdam (GRAPPA), University of Amsterdam, Science Park 904, NL-1098 XH Amsterdam, the Netherlands*

## ABSTRACT

Type Ia supernovae (SNe Ia), standardizable candles that allow tracing the expansion history of the Universe, are instrumental in constraining cosmological parameters, particularly dark energy. State-of-the-art likelihood-based analyses scale poorly to future large data sets, are limited to simplified probabilistic descriptions, and must explicitly sample a high-dimensional latent posterior to infer the few parameters of interest, which makes them inefficient. Marginal likelihood-free inference, on the other hand, is based on forward simulations of data, and thus can fully account for complicated redshift uncertainties, contamination from non-SN Ia sources, selection effects, and a realistic instrumental model. All latent parameters, including instrumental and survey-related ones, per object and population-level properties, are implicitly marginalized, while the cosmological parameters of interest are inferred directly. As a proof of concept, we apply truncated marginal neural ratio estimation (TMNRE), a form of marginal likelihood-free inference, to BAHAMAS, a Bayesian hierarchical model for SALT parameters. We verify that TMNRE produces unbiased and precise posteriors for cosmological parameters from up to 100 000 SNe Ia. With minimal additional effort, we train a network to infer simultaneously the ∼100 000 latent parameters of the supernovae (e.g. absolute brightnesses). In addition, we describe and apply a procedure that utilizes local amortization of the inference to convert the approximate Bayesian posteriors into frequentist confidence regions with exact coverage. Finally, we discuss the planned improvements to the model that are enabled by using a likelihood-free inference framework, like selection effects and non-Ia contamination.

**Key words:** (cosmology:) cosmological parameters – methods: statistical.

## 1 INTRODUCTION

The momentous discovery of the accelerated expansion of the universe (Riess et al. 1998; Perlmutter et al. 1999) demonstrated that type Ia supernovae (SNe Ia) can be used as standardizable candles to measure extragalactic distances, which, together with a determination of redshift, lead to constraints on the cosmological parameters. SNe Ia are also useful to extend the distance ladder, thus allowing, when properly calibrated, to measure the Hubble–Lemaître constant, $H_0$ – a measurement famously and mysteriously in tension with the result obtained from cosmic microwave background (CMB) data at high redshift (Di Valentino et al. 2021).

While the initial discovery of cosmic acceleration relied on a total of 58 high-redshift SNe Ia (backed by a similar-sized low-redshift anchor sample), several dedicated SN Ia campaigns: CfA 1–4 (Perlmutter et al. 1999; Jha et al. 2006; Hicken et al. 2009, 2012), BSNIP (Silverman et al. 2012), SNLS (Guy et al. 2010), CSP I–II (Krisciunas et al. 2017; Phillips et al. 2019), general transient searches: Pan-STARRS 1 (Scolnic et al. 2018), Foundation (Foley et al. 2018), and extensive cosmological surveys: SDSS (Sako et al. 2018), DES (Brout et al. 2019), have since enlarged the spectroscopically confirmed cosmological sample to about 2000 SNe Ia (Scolnic et al. 2022). Similarly to the Universe, the rate of SN Ia discovery is accelerating exponentially, and in the near future, the Vera Rubin Observatory will observe ∼$10^5$ SNe Ia per year in its Legacy Survey of Space and Time (LSST; LSST Science Collaboration 2009; Ivezić et al. 2019).

This massive increase in the quantity of the data has spurred the development of increasingly sophisticated statistical methods for their analysis, in order to accurately model the data collection procedure in all of its complexity, to capture important systematic effects, and to describe the underlying population variability of both SNe Ia and host-galaxy properties. These efforts have become central in ensuring more precise and accurate constraints on the dark energy redshift evolution from future large samples of SNe Ia.

Traditionally, cosmological inference with type Ia SNe has relied on *standardization*: the process of *correcting* their observed brightnesses so that when the dimming effect of distance is taken into account, the SN Ia population is as homogeneous as possible. Initially, this was done using hand-crafted descriptions of the observed light curves (Pskovskii 1967, 1977, 1984; Phillips 1993; Perlmutter et al. 1997), which were later replaced by summaries derived from data. By far the most popular such model is SALT (Guy et al. 2005, 2007; Betoule et al. 2014; Kenworthy et al. 2021; Taylor et al. 2021), which decomposes the SN Ia spectral energy distribution (SED) using functional principal component analysis: a methodology extended in SNEMO (Saunders et al. 2018) to up to 15 components. A particular

★ E-mail: kkarchev@sissa.it

SN Ia is then described by the relative contributions of the principal components and by a colour parameter, which are subsequently used for standardization and state-of-the-art cosmological inference from SNe Ia (e.g. Abbott et al. 2019; Brout et al. 2022).

Following the realization that the observed excess variability of SNe Ia after standardization is not simply described by additional Gaussian noise/scatter, in the last decade, sophisticated Bayesian hierarchical models (BHMs) were developed. Some, like BAHAMAS (March et al. 2011; Shariff et al. 2016b; Rahman et al. 2022), UNITY (Rubin et al. 2015), the model of Ma, Corasaniti & Bassett (2016), simple-BayeSN (Mandel et al. 2017), and Steve (Hinton et al. 2019), describe SNe Ia using summary parameters obtained from SALT fits but fundamentally distinguish observational noise from intrinsic variability. In contrast, Mandel et al. (2009, 2022) and Mandel, Narayan & Kirshner (2011) presented a fully Bayesian treatment of the underlying SN Ia SED template in a model called BayeSN, which was recently used for a cosmological analysis of the RAISIN sample (Jones et al. 2022).

Hierarchical Bayesian modelling, however, comes at the cost of having to infer individual parameters for each observed SN Ia, and so the computational burden scales with the data set size. While for current surveys and compilations of $N \sim 1000$ SNe Ia this is feasible with methods like Hamiltonian Monte Carlo (HMC) and Gibbs sampling (provided the conditional independence of global parameters is not negated by e.g. selection effects), such methods will not be applicable to larger data sets, e.g. due to systematic uncertainties which necessitate the inversion of large dense matrices, incurring an $\mathcal{O}(N^3)$ penalty even in the case of a Gaussian likelihood. Moreover, whereas light-curve-summarizing techniques provide 1– 10 parameters per object, modelling SEDs as in BayeSN requires $\sim 1000$ parameters for each supernova, and so joint inference quickly becomes impossible (e.g. Jones et al. 2022 and Mandel et al. 2022 analyse only 79 SNe Ia).

Even if computation were not an issue, all likelihood-based analyses share one fundamental limitation: they need to be explicit in their probabilistic description of the data, including 'physical' (intrinsic) properties of the supernovae and the environment (e.g. dust), their population-level distributions (which possibly evolve through time), instrumental effects and survey cross-calibration, the observational selection process, and purity of the sample. This results in necessarily simplified descriptions, consisting predominantly of normal distributions (out of computational convenience), and, where probabilities are intractable (or hard to compute, e.g. for selection effects), in non-principled or ad hoc de-biasing procedures. As we demonstrate in Section 4, in some cases, simplifications required to render the model tractable (e.g. propagating photometric redshift uncertainties linearly on to magnitudes) lead to biases in cosmological inference that only become apparent with large data samples.

On the other hand, the major modelling challenges – intrinsic variability, dust properties, redshift uncertainty from photo-z, selection effects, and contamination – are all relatively straightforward to simulate. Indeed, some recent analyses already employ forward simulations with the comprehensive SNANA package (Kessler et al. 2009a) for parts of their workflow, e.g. for calibration (Burke et al. 2018), inferring the distribution of dust properties (Popovic et al. 2021a) and possible correlations with features of the host (Popovic et al. 2021b), and, most prominently, correcting for Malmquist bias (Malmquist 1922, 1925) in the derived distance moduli (see e.g. Kessler & Scolnic 2017). These studies use forms of approximate Bayesian computation (ABC; see e.g. Sisson, Fan & Beaumont 2018), an established technique belonging to the class of likelihood-free simulation-based inference (LF-SBI), which was also used directly for cosmological inference from SNe Ia by Weyant, Schafer & Wood-Vasey (2013) and Jennings, Wolf & Sako (2016). ABC, however, requires hand-crafted distance measures comparing simulation output to the data or to (hand-crafted) summary representations and an abundance of simulations to produce accurate results.

Together with advances in computer science like automatic differentiation and more straightforward graphics processing unit (GPU) utilization, which enabled the widespread adoption of neural networks (NNs), recent years have seen the introduction of new NN-based likelihood-free Bayesian analysis techniques (see Cranmer, Brehmer & Louppe 2020; Lueckmann et al. 2021, for reviews). In these methods, a stochastic simulator is used as an implicit representation of the likelihood, whose numerical evaluation is no longer necessary. This simplifies the inference procedure and shifts the focus to constructing a realistic simulator: an often much easier feat than reverse-solving a Bayesian model. Importantly, it allows the simultaneous inclusion of all relevant processes influencing the data, including those like selection effects, for which a probabilistic description is either intractable or computationally prohibitive.

This work is meant as a first proof of concept that a likelihood-free methodology can deliver accurate and precise posteriors for cosmological parameters from up to $\sim 10^5$ SNe Ia. We only use summary statistics derivable with SALT in the context of an intentionally simple BHM resembling BAHAMAS so as to enable verification of the likelihood-free posteriors on mock data. We discuss our immediate plans towards a realistic simulator in Section 2.6.

We employ truncated marginal neural ratio estimation (TMNRE; Miller et al. 2020, 2022), a sequential implementation of the general neural ratio estimation (NRE) technique (Hermans, Begy & Louppe 2020) that composes well with marginalization. It converts posterior inference into a classification problem and then solves it by training a neural network on simulated data. The key to making the analysis scalable even in the presence of numerous parameters describing the individual SNe Ia is that TMNRE only targets low-dimensional (one- or two-dimensional in this work) *marginal* posteriors for the parameters of interest, rather than the *joint*, high-dimensional posterior as necessary in likelihood-based inference [including with Markov chain Monte Carlo (MCMC) and variational simulation-based inference (SBI) techniques]. In addition, TMNRE iteratively refines the regions in parameter space from which training examples are drawn based on a given target observation, so as to maximize the simulator efficiency and utilization of the network's learning capacity.

This paper, a first example of Supernova Ia Cosmology with truncated marginal neural Ratio EsTimation (SICRET), is structured as follows. We present in Section 2 the Bayesian hierarchical model we use in this proof-of-concept paper and discuss the ways we plan to improve it in Section 2.6. In Section 3, we elaborate TMNRE, including the truncation scheme used to zoom into regions of high posterior density, and the network architecture we adopt. We demonstrate the inference procedure on mock data in Section 4. In Section 4.1, we infer marginally the cosmological parameters and show that TMNRE can derive accurate posteriors from $10^5$ SNe Ia utilizing the full model complexity, whereas model simplifications like linear uncertainty propagation, necessary for MCMC sampling, may introduce a significant bias. In Section 4.2, we show calibrated (i.e. with exact coverage) confidence regions for the cosmological parameters, derived using a procedure presented in Section 3.4 and Appendix A. We also verify, in Section 4.3, that the TMNRE posteriors are precise, i.e. the inference procedure is able to combine information from the full data set. Lastly, in Section 4.4, we perform simultaneous marginal inference of $10^5$ latent parameters of the supernovae. We discuss our results and present our conclusions in Section 5.

## 2 MODEL: BAHAMAS

In this section, we present in detail the forward model used for our analysis. It is implemented as a simulator that generates mock SN Ia data, which is then used to train an inference network, as described in Section 3. The simulator consists of layers: *global* parameters control the distributions of *latent* variables, which describe individual SNe Ia; finally, an instrument model produces observables (*data*) from the latent parameters.

In this proof-of-concept paper, we make a number of simplifying assumptions, which will be relaxed in future works in order to apply TMNRE to real data. We discuss them in Section 2.6.

### 2.1 Observables

We focus on inference from *processed* multiband SN Ia light curves that are summarized in the *observed* SALT parameters for each supernova $s$:

$$\boldsymbol{d}^s = [\hat{m}^s, \hat{x}_1^s, \hat{c}^s], \quad (1)$$

where $x_1$ and $c$ are parameters describing the light-curve shape (its 'stretch' and 'colour', respectively), and $m$ is the peak apparent magnitude of the supernova. For $N$ observed SNe Ia this is a length-$(N \times 3)$ vector, $\boldsymbol{d}$, for which analysis pipelines usually provide an *observational covariance* matrix $\hat{\boldsymbol{\Sigma}}$ of size $3N \times 3N$, which could include correlations between different SNe Ia due to systematic effects and uncertainties in the SALT model. The sampling distribution of the data is, therefore, assumed Gaussian:

$$\boldsymbol{d} \sim \mathcal{N}([\boldsymbol{m}, \boldsymbol{x_1}, \boldsymbol{c}], \hat{\boldsymbol{\Sigma}}). \quad (2)$$

Here $m^s$, $x_1^s$, $c^s$ are the *latent* parameter values of a SN Ia, of which noisy estimates $\hat{m}^s$, $\hat{x}_1^s$, $\hat{c}^s$ are observed.

Finally, each supernova has a latent cosmological redshift $z^s$, of which a measurement $\hat{z}^s$ is made. We will not forward-simulate the redshift measurements; for that we will need a prior, related to SN Ia rates, as we discuss in Section 2.6.3. Instead, we will assume that the observations from which $\hat{z}^s$ has been determined (e.g. multiband host photometry) are independent of the SN Ia light-curve data and result in a *posterior* for the latent redshift, whose form follows the toy model in Roberts et al. (2017):

$$z^s \mid \hat{z}^s \sim \mathcal{N}\left(\hat{z}^s, (1+\hat{z}^s)^2 \sigma_z^2\right), \quad (3)$$

where $\sigma_z$ is a global parameter (to be inferred) controlling the overall redshift variance. We will use this posterior as a *prior* and forward-simulate the latent redshift from it.

### 2.2 Supernova model

The 'physical' model of SN Ia latent variables includes two components. Firstly, the light-curve parameters are linearly correlated with the supernova's intrinsic magnitude, $M^s$, via the so-called Tripp formula (Phillips 1993; Tripp 1997, 1998):

$$M^s = M_0^s - \alpha x_1^s + \beta c^s, \quad (4)$$

where $\alpha$ and $\beta$ are the correction coefficients, global to all SNe Ia, and $M_0^s$ is the supernova's absolute magnitude *post-corrections*.[1]

### 2.3 Cosmology

The second part of the model is the influence of cosmology, which converts between absolute and apparent magnitudes:[2]

$$m^s = M^s + \mu(\mathcal{C}, z^s) \quad (5)$$

with $\mu^s \equiv \mu(\mathcal{C}, z^s)$ the distance modulus of a supernova with cosmological redshift $z^s$ under a cosmological model labelled with $\mathcal{C}$. We will assume radiationless Lambda cold dark matter ($\Lambda$CDM), parametrized by $\mathcal{C} \equiv \{\Omega_{m0}, \Omega_{\Lambda 0}\}$: the present-day relative densities of matter and dark energy, respectively; however, our method is directly applicable to any cosmological model for which the distance modulus can be calculated.[3]

### 2.4 Prior model

We build a Bayesian hierarchical model for supernova cosmology *à la* BAHAMAS (Shariff et al. 2016a, b) and assume normal priors on $M_0^s$, $x_1^s$, and $c^s$. The respective prior means ($\bar{M}_0$, $\bar{x}_1$, $\bar{c}$) and standard deviations[4] ($\sigma_{res}$, $R_{x_1}$, $R_c$) are in turn global *hyperparameters* for which (hyper)priors are assumed: normal for the means, an inverse-gamma distribution for the residual scatter, and log-normal for the stretch and colour variances. For the redshift uncertainty, $\sigma_z^2$, we also take an appropriate inverse-gamma prior. All SN Ia-related parameters, along with their priors, their hyperparameters, and their hyperpriors are listed in Table 1. Finally, for the parameters of $\Lambda$CDM we use a uniform prior over $[0; 2]^{\otimes 2}$ in $\Omega_{m0}$–$\Omega_{\Lambda 0}$ space and use $\Omega_{m0} = 0.3$ and $\Omega_{\Lambda 0} = 0.7$ for generating mock data. The full model is depicted as a directed acyclic graph in Fig. 1.

### 2.5 Simulator configuration: mimicking Pantheon

The simulator requires two fixed inputs: the vector of observed redshifts, $\hat{\boldsymbol{z}}$, and the observational covariance, $\hat{\boldsymbol{\Sigma}}$. We build these based on the Pantheon compilation (Scolnic et al. 2018), so that the SNe Ia we simulate have a realistic distribution of redshifts[5] and reasonably sized observational uncertainties, appropriate for the supernova's redshift. We select $N$ supernovae from Pantheon (with replacement when $N$ is larger than the compilation size)[6] and concatenate their reported redshifts into $\hat{z}$ and stack diagonally their *individual* parameter covariance matrices (each of size 3 × 3) to form a block-diagonal $\hat{\Sigma}$.

---

[1]The term 'correction' comes from the tradition of SN Ia standardization, which is a technique of reverse modelling. From a forward-modelling perspective, $M_0^s$ is a 'noisy' realization (i.e. with scatter) of the brightness of the standard SN Ia ($x_1 = 0$, $c = 0$).

[2]This equation, strictly, applies only to bolometric magnitudes; otherwise, one needs to account for the redshifting of light in and out of the observed band, which is usually achieved through so-called *K-corrections*. Indeed, SN Ia magnitudes ($m$) are usually reported in the *rest-frame B*-band, and all *K*-correction has already been applied during the analysis of light curves.

[3]Regardless of the particular cosmological model, the luminosity distance, to which the distance modulus is directly related, can be expanded in a Taylor series, whose coefficients can be regarded as the 'cosmological parameters of interest'. To second order, the expansion is (Weinberg 2008, equation 1.4.9) $d_L = H_0^{-1} \left[ z + \frac{1}{2}(1-q_0)z^2 + \mathcal{O}(z^3) \right]$, where $H_0$ is the Hubble–Lemaître constant, and $q_0$ is the *deceleration parameter*. Evidently, $H_0^{-1}$ simply sets the absolute distance scale (i.e. the distance units) and is therefore degenerate with the absolute brightness scale of the SNe Ia (i.e. the luminosity units). Thus, the 'first-order' effect that can be derived from SN Ia data is that of $q_0$. Indeed, in the posteriors presented in Section 4, the best constrained direction is that corresponding to varying $q_0 = \Omega_{m0}/2 - \Omega_{\Lambda 0}$ in $\Lambda$CDM (Peebles 1993, equation 13.7).

[4]The interpretation of the so-called *residual scatter*, $\sigma_{res}$, is multifold: it could be taken to represent a *scatter* of the absolute magnitudes of SNe Ia or equivalently an additional unaccounted-for observational *noise* (hence the label: *residual* after standardization).

**Table 1.** SN Ia parameters and priors and hyperparameters and hyperpriors and 'true' values used to generate mock data. See also Fig. 1 for a graphical representation of the model.

| Parameter | | (Hyper)prior | Mock value |
|---|---|---|---|
| Latent redshift | $z^s$ | $\mathcal{N}(\hat{z}^s, (1+\hat{z}^s)^2 \sigma_z^2)$ | |
| Measured redshift | $\hat{z}^s$ | | |
| Redshift uncertainty | $\sigma_z^2$ | $\gamma^{-1}(0.0003, 0.0003)$ | $0.04^2$ |
| Correction coefficients | $\alpha$ | $\mathcal{U}(0, 1)$ | 0.14 |
| | $\beta$ | $\mathcal{U}(0, 4)$ | 3.1 |
| Abs. magnitude | $M_0^s$ | $\mathcal{N}(\bar{M}_0, \sigma_{\text{res}}^2)$ | |
| Mean abs. mag. | $\bar{M}_0$ | $\mathcal{N}(-19.3, 2^2)$ | $-19.5$ |
| Residual mag. scatter | $\sigma_{\text{res}}^2$ | $\gamma^{-1}(0.003, 0.003)$ | $0.1^2$ |
| 'Stretch' covariate | $x_1^s$ | $\mathcal{N}(\bar{x}_1, R_{x_1}^2)$ | |
| $x_1$ Prior mean | $\bar{x}_1$ | $\mathcal{N}(0, 10^2)$ | 0 |
| $x_1$ Prior st. dev. | $R_{x_1}$ | $\log\mathcal{U}(10^{-5}, 10^2)$ | 1 |
| 'Colour' covariate | $c^s$ | $\mathcal{N}(\bar{c}, R_c^2)$ | |
| $c$ Prior mean | $\bar{c}$ | $\mathcal{N}(0, 1^2)$ | 0 |
| $c$ Prior st. dev. | $R_c$ | $\log\mathcal{U}(10^{-5}, 10^2)$ | 0.1 |

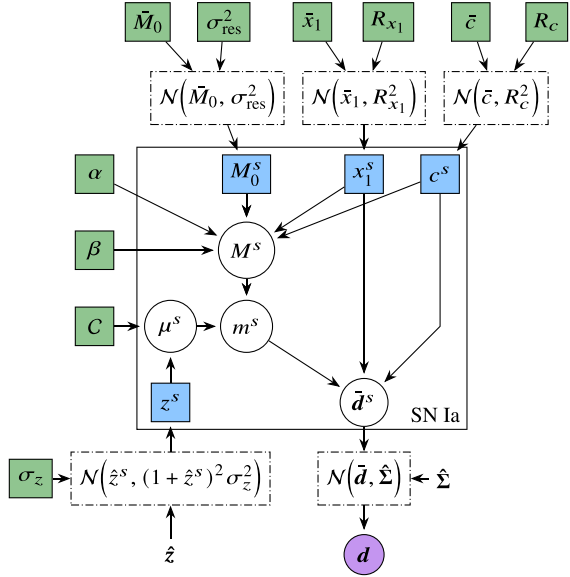

**Figure 1.** Graphical representation of the model. Filled squares indicate parameters: *global* outside and *latent* inside the SN Ia plate, which indicates the conditional independence a priori of the SNe Ia (each labelled with 's') from one another. Unfilled circles represent deterministic nodes, while the filled circle is the observed data. $\hat{z}$ and $\hat{\Sigma}$ are fixed inputs to the model.

---

[5]We note that future large surveys will have a different distribution of $\hat{z}$ than Pantheon, since their selection probabilities will be different (see e.g. LSST Science Collaboration 2009, fig. 11.1, for the expectations for the Vera Rubin Observatory). And since the statistical power of a SN Ia sample for cosmological parameter inference depends strongly on this distribution, our posteriors from $10^5$ SNe Ia should be taken as an indication rather than prediction of possible future constraints.

Cosmological analyses usually consider additionally dense statistical and systematic covariances (see e.g. Conley et al. 2011), representing, respectively, uncertainties in the underlying SN Ia model (SALT) and the effect of e.g. instrument calibration. While the latter can be efficiently and rigorously replaced in an SBI framework by forward-simulating the relevant systematics, model uncertainties can be propagated only by performing fits to light curves. Furthermore, having a dense $\hat{\Sigma}$ would affect the performance of the simulator, in its current form, by introducing an $\mathcal{O}(N^2)$ computational complexity – and a corresponding memory footprint – when sampling $N$ mock SNe Ia from latent variables. The expensive Cholesky decomposition which this requires, however, only needs to be performed once at the beginning of the analysis if $\hat{\Sigma}$ is fixed, i.e. instrumental and SALT model parameters are not being varied. In this proof of concept we choose not to go beyond the block-diagonal data covariance and discuss in Section 2.6.1 how we plan to address these issues more naturally as part of future realistic forward simulators.

### 2.6 Room for improvement

Improvements to the model which need to be addressed before it can be used for analysis of real data fall in several broad categories: regarding the nature of the analysed data, the effect of the SN Ia environment (properties of the host galaxy), measuring its redshift, object classification, and sample selection. Most of these are unwieldy in a likelihood-based framework either due to the presence of complicated distributions or their outright intractability. On the other hand, as long as they can be included in a forward simulator, none of these modifications impose fundamental changes to a likelihood-free inference (LFI) procedure. Specifically, TMNRE of the cosmological parameters is not affected in any way.

#### 2.6.1 Intrinsic SN Ia model and observed data

Raw SN Ia data is (exclusively, for the vast majority of SNe Ia expected from large future surveys) in the form of light curves: a collection of flux observations at different times and in different broad spectral bands, and, to a lesser extent, spectra. To properly analyse them, a model needs a representation of the *intrinsic* SED of the supernova. On the other hand, SALT parameters only summarize *observed* properties of the light curves (and spectra), and it is clear that describing the intrinsic SED with just the *latent* SALT parameters does not capture its full complexity and the observed variations within the SN Ia population (Saunders et al. 2018; Mandel et al. 2022). Moreover, it has been found that the SN Ia population itself evolves with cosmic time and that this can bias cosmological inference (Lee et al. 2020; Nicolas et al. 2021). This effect can be included in a BHM by allowing the global parameters ($\bar{M}_0$, $\sigma_{\text{res}}^2$, etc.) to be functions of redshift, represented e.g. as splines whose parameters are also inferred (see e.g. Rubin et al. 2015).

Likelihood-based analyses rely on simple distributions and empirical prescriptions to capture these effects, whereas a forward simulator can more explicitly describe the underlying SEDs, while avoiding the need to calculate a covariance matrix for the summary parameters and instead fully accounting for the SED uncertainties. Finally, shifting the analysis towards raw flux measurements allows principled modelling of all instrumental effects: Poisson-distributed observed photon counts, background and instrumental noise, light from the host, and zero-point calibration, all of which are currently being summarized in a simple Gaussian systematic uncertainty on the corrected magnitudes. In contrast, a forward simulation of data acquisition can be arbitrarily complex.

---

[6]Nominally, the Pantheon data release contains 1048 SNe Ia, but for two of them ('16232' and 'PTF10bjs') the reported parameter covariances are not positive definite... so we only use 1046.

### 2.6.2 Environment

A well-known effect that is difficult to include on the level of summary parameters, but easily applied to SED-based light-curve models, is that of the SN Ia environment, commonly referred to as 'dust extinction', which is dictated by other properties of the host galaxy like its mass and star-formation rate (SFR; Brout & Scolnic 2021). In that respect, the model also needs to account for variations in the extinction properties among and within host galaxies. Extinction within the Milky Way (Schlafly & Finkbeiner 2011), which depends on the location of the SN Ia on the sky, needs to be taken into account as well. It should be noted that extinction is not the only 'effect' that host properties have on a supernovae: it is possible that type Ia SNe that come from different stellar populations have different intrinsic properties (Childress et al. 2013). The locations of SNe Ia within their hosts have also been shown to be important (Hill et al. 2018), although there is little hope in having constraints for them in large surveys.

A forward simulator enables correlations between arbitrary host and SN Ia properties to be included under the control of specialized parameters, in a manner similar to the Tripp formula, while still allowing for additional effects to be included as 'residual scatter'. Together with a forward model for host observations based on underlying parameters like mass, SFR, etc., this opens the door to general simultaneous analysis of supernovae and their hosts.

### 2.6.3 Redshift

As we will show in Section 4, proper treatment of the redshift of a SN Ia is crucial for obtaining unbiased cosmological posteriors, especially in the era of large-scale surveys that will discover far too many SNe Ia for spectroscopic follow-up. One will need to rely, then, on estimates derived from multiband host photometry, which often exhibit departures from normality and even multimodality (see e.g. Leistedt, Mortlock & Peiris 2016; Linder & Mitra 2019), and so the Gaussian approximation in equation (3) will not be adequate. Apart from using a complicated likelihood, one must also place a redshift prior based on SN Ia rates, which come with their own set of parameters and are related to the cosmological model.

Moreover, in this work, we assume that the *cosmological* redshift (which enters in the distance modulus calculation) is directly observable. In reality, the observed light is affected also by a variety of peculiar motions: of the host galaxy with respect to the CMB frame, of the SN Ia itself inside the galaxy, and, similarly, of the Milky Way and Solar System (and Earth).[7] Whereas there are stringent constraints on the latter two from CMB anisotropies (Planck Collaboration I 2020), the motions of far-away galaxies are known only on average, e.g. from bulk galaxy-flow models (Boruah, Hudson & Lavaux 2020, 2021), whose intrinsic scatter must be taken into account. Similarly to its position, the peculiar velocity of a SN Ia within its host is hardly observable and, therefore, a source of additional peculiar-velocity scatter.

Failing to model peculiar velocities has been shown to cause a bias to cosmological inference when the sample contains low-redshift SNe Ia (Wojtak et al. 2015; Linder & Mitra 2019; Huterer 2020). While early studies, therefore, *discarded* SNe Ia with e.g. $z \lesssim 0.02$ (Kessler et al. 2009b), more recent analyses usually correct redshifts to the CMB reference frame and add the various *spreads* in quadrature to form an *uncertainty*, or resort to high-dimensional Monte Carlo techniques (e.g. Gibbs or Hamiltonian MC; Rahman et al. 2022). LF-SBI provides an explicit way to include all sources of redshift (and the correlations they induce between different SNe Ia) without the need to explicitly infer $\mathcal{O}(N)$ additional parameters.

### 2.6.4 Contamination

Without spectroscopic data for the supernova, one must consider the possibility of two types of contamination, namely, the presence of non-SN Ia events in the sample, and misidentification of the host, from which the redshift is derived. Unaccounted for, any of these effects can introduce a significant bias to the inference of cosmological parameters (Roberts et al. 2017).

Contamination can be dealt with by explicitly calculating the odds ratio of an event being a SN Ia or a contaminant (Kunz, Bassett & Hlozek 2007). A likelihood-free framework learns this implicitly by observing that in simulations certain measurements (interpreted as contamination) do not contribute information to the cosmological inference task, or, if they do, in what way they differ from type Ia supernovae. While SBI still relies on having an explicit model for the possible contaminants, these can be arbitrarily complex. Going beyond the simple assumption of a Gaussian dispersion of absolute magnitudes used in past analyses (Hlozek et al. 2012; Jones et al. 2017), one can, e.g. use any non-Ia template from SNANA (Kessler et al. 2009a) or sophisticated data-driven models (Revsbech, Trotta & van Dyk 2018; Boone 2019), setting a priori probabilities on each object's class during the forward run.

### 2.6.5 Selection effects

Finally, an assumption implicit in our hierarchical model is that the population-level distributions (i.e. the priors on the latent variables $M_0$, $x_1$, $c$, and $z$) are appropriate for *observed*, as opposed to *all* SNe Ia in the Universe. That is, all probabilities in our model are implicitly conditional on the SN having been observed and selected for inclusion in the data set. If one assumes the converse – that population distributions indeed reflect the whole population – one needs to deal with *selection effects*, which in general modify the shape *and normalization* of downstream distributions in ways that depend on the cosmological and other global parameters and usually require numerical or Monte Carlo integration in order to calculate the total selection probability at each step of a sampling chain. On the other hand, a simulator used for likelihood-free inference only needs to faithfully recreate the survey strategy and various quality cuts applied during the acquisition and reduction of the real data.

---

[7]Papers in the literature have even considered the velocity of ejecta in a SN Ia explosion (Foley 2012) and a gravitational redshift due to density fluctuations (Davis et al. 2011; Wojtak, Davis & Wiis 2015; Calcino & Davis 2017).

# 3 INFERENCE: TRUNCATED MARGINAL NEURAL RATIO ESTIMATION

## 3.1 Marginal ratio estimation

We aim to infer from data $d$ a subset $\Theta$ of the parameters of a hierarchical Bayesian model, i.e. derive the posterior

$$p(\Theta \mid d) = \frac{\int p(d \mid \Theta, Z) p(\Theta, Z) \, dZ}{p(d)} \qquad (6)$$

marginalized over the nuisance[8] parameters $Z$. Often, the integral, which describes the marginal likelihood of the data given *only* the parameters of interest, cannot be calculated due to the high dimensionality of $Z$ or because some of the distributions involved are inherently unknown. However, if the model can be implemented as a forward simulator, i.e. if samples from the distributions involved can be drawn, the posterior of interest can be obtained via (neural) estimation of the likelihood-to-evidence ratio, as we now describe.

Ratio estimation reformulates Bayesian inference as a classification task, underpinned by two realizations. First, that a Bayes optimal classifier (i.e. which minimizes the Bayesian risk of misclassification) trained to distinguish samples from two distributions $p_1(x)$ and $p_2(x)$ must consider the ratio of their probability densities, $p_1(x)/p_2(x)$ (Devroye, Györfi & Lugosi 1996). Thus, even if the densities $p_1(x)$ and $p_2(x)$ themselves are not known, one can approximate $p_1(x)/p_2(x)$ by training a classifier on samples $x_{1,i} \sim p_1(x)$ and $x_{2,i} \sim p_2(x)$.

The second realization is that Bayes' theorem equates three ratios of probability densities: the likelihood $p(d \mid \Theta)$ to evidence $p(d)$, the posterior $p(\Theta \mid d)$ to prior $p(\Theta)$, and the joint $p(\Theta, d)$ to product of marginals $p(\Theta) p(d)$:

$$r(\Theta, d) \equiv \frac{p(\Theta, d)}{p(\Theta) p(d)} = \frac{p(d \mid \Theta)}{p(d)} = \frac{p(\Theta \mid d)}{p(\Theta)}, \qquad (7)$$

which means that one can gain access to the posterior-to-prior ratio (and subsequently use it for inference) by training a classifier of joint versus marginal pairs, which are easily obtainable with a forward simulator.

Concretely, joint pairs $(\Theta_i, d_i) \sim p(\Theta, d) \equiv p_1(\Theta, d)$ are generated by first sampling all parameters from the (hierarchical) prior $(\Theta, Z \sim p(\Theta, Z))$ and then generating mock observations with the simulator conditioned on the sampled values. Ignoring the values of nuisance parameters from this point forward is the SBI equivalent of the marginalization in equation (6). On the other hand, the sets of samples $\{\Theta_i\}$ and $\{d_i\}$, taken separately, are distributed according to the respective marginals: the (marginal) prior $p(\Theta)$ and the evidence $p(d)$, and so random pairing results in samples $(\Theta_i, d_j) \sim p(\Theta) p(d) \equiv p_2(\Theta, d)$.

As the last equality in equation (7) suggests, if the prior is tractable, $r(\Theta, d)$ gives direct access to the posterior density: $p(\Theta \mid d) = r(\Theta, d) p(\Theta)$. Alternatively, one can use $r(\Theta, d)$ to weight prior samples, so that they are re-interpreted as such from the posterior. The priors used in this work are simple and analytic, and we will, therefore, be evaluating and analysing the marginal posteriors (which are at most 2D) on regular grids spanning the prior support in the respective parameter spaces, with the notable exception of the latent parameters, for which we use re-weighted samples, as discussed in Section 4.4.

Ratio estimation as a Bayesian inference technique does not require actual classification of pairs $(\Theta, d)$, i.e. assigning concrete labels to examples. Instead, the classifier is constructed in such a way as to make an estimate of $r(\Theta, d)$ explicitly calculable (see below). In a typical classification task, the ratio will then be used to calculate the class probabilities: $r/(1 + r) = p_1/(p_1 + p_2)$ and $1/(1 + r) = p_2/(p_1 + p_2)$, based on which an example will be classified.

## 3.2 Neural ratio estimation

In the practice of *neural* ratio estimation (NRE), as proposed by Hermans et al. (2020), the classifier is a neural network (NN), dubbed the *inference network*, that takes as input a parameters-of-interest–data pair and produces an estimate of the joint-to-marginal ratio: $\hat{r}_\Phi(\Theta, d)$, or, for numerical reasons, its logarithm $\ln \hat{r}_\Phi(\Theta, d)$. The network parameters $\Phi$ are optimized via stochastic gradient descent to minimize the binary cross-entropy:

$$\mathcal{L}(\Phi) = \mathbb{E}_{p(\Theta,d)} \left[ -\ln \frac{\hat{r}_\Phi(\Theta, d)}{1 + \hat{r}_\Phi(\Theta, d)} \right]$$
$$+ \mathbb{E}_{p(\Theta) p(d)} \left[ -\ln \frac{1}{1 + \hat{r}_\Phi(\Theta, d)} \right], \qquad (8)$$

which one can verify indeed results in $\hat{r}_\Phi(\Theta, d) \to r(\Theta, d)$.

During training, we estimate the loss using two joint pairs, $(\Theta_1, d_1)$ and $(\Theta_2, d_2)$, sampled independently of one another, intermingling them to produce marginal pairs $(\Theta_1, d_2)$ and $(\Theta_2, d_1)$. Then, we perform the gradient descent on the loss

$$2l(\Phi) \approx -[\ln \sigma(\ln \hat{r}_\Phi(\Theta_1, d_1)) + \ln \sigma(-\ln \hat{r}_\Phi(\Theta_2, d_1))]$$
$$- [\ln \sigma(\ln \hat{r}_\Phi(\Theta_2, d_2)) + \ln \sigma(-\ln \hat{r}_\Phi(\Theta_1, d_2))], \qquad (9)$$

where $\sigma(x) \equiv [1 + \exp(-x)]^{-1}$ is the sigmoid function. We exploit the simulator to continually produce training samples for online learning, i.e. we do not have fixed training, validation, etc. sets.

### 3.2.1 Parameter groups

In scientific applications, one is often interested in a number of 1- or 2-dimensional marginal posteriors, so we might have multiple parameter subsets (groups) $\Theta$ that we want to consider separately. We will differentiate groups of *global* parameters, labelling them with $\theta_i$, from those consisting of a single object's *latent* parameter(s) of interest: $\vartheta_i^s$. Usually, each parameter of interest will figure in exactly one group, but this is not required. The posteriors that we are interested in are, then, the set $\{p(\theta_i \mid d)\}_i \cup \{p(\vartheta_i^s \mid d)\}_{i,s}$.

Concretely, when we are inferring all global parameters, we will have *10* 'groups': $\{\theta_i\} \equiv \{[\Omega_{m0}, \Omega_{\Lambda 0}], \sigma_z, \alpha, \beta, \bar{M}_0, \sigma_{\text{res}}, \bar{x}_1, \log_{10} R_{x_1}, \bar{c}, \log_{10} R_c\}$, one of which is 2-dimensional, and when we are inferring cosmology only, we will have only $\theta_1 \equiv [\Omega_{m0}, \Omega_{\Lambda 0}]$. Finally, we will eventually have $\vartheta_1^s \equiv M_0^s$.

### 3.2.2 Network architecture

Learning the parameter groups marginally means having separate ratio estimators particular to each of them. This motivates a structure for the inference network split into a data pre-processor (*head*) and multiple *tails*, containing a parameter pre-processor and a ratio estimator. At each step of training, we pre-processed the data once and pass it as input to all tails. Then we evaluate equation (9) with each parameter group $\Theta$ independently and combine the obtained losses before the gradient descent step. This means that the head network is updated based on the performance across all parameters of interest, so it needs to extract relevant information for all of them.

---

[8] Signifying, here, any hierarchic parameters that are deemed not of interest.

Below, we elaborate on the different components of the inference network. We list the particular implementations in Table 2 and depict the network graphically in Fig. 2. We will omit for clarity the subscripts indicating dependence on network parameters, jointly labelled **Φ** before.

The head network operates independently on the observations from each supernova and produces a number of non-linear *features* for it:

$$\boldsymbol{d}^s = \mathtt{DataHead}(\boldsymbol{d}^s), \tag{10}$$

which it then combines into a number of *summary statistics* that describe the data set as a whole:

$$\mathcal{S} = \mathtt{Summariser}([\boldsymbol{d}^s]_{s=1}^N). \tag{11}$$

The summariser encodes information on the statistics of the data set as a whole, which relate to the global model parameters. It is, therefore, important to preserve any ordering of the data, in our case stemming from the measured redshifts and observational uncertainties, while simulating training data (see also Section 3.2.3).

**Each tail network** first featurizes the parameter(s) of interest that it is responsible for:

$$\begin{aligned} \text{global:} \ \boldsymbol{\theta}_i &= \mathtt{ParamHead}_{\boldsymbol{\theta}_i}(\boldsymbol{\theta}_i), \\ \text{latent:} \ \boldsymbol{\vartheta}_i^s &= \mathtt{ParamHead}_{\boldsymbol{\vartheta}_i}(\boldsymbol{\vartheta}_i^s), \end{aligned} \tag{12}$$

thus possibly accounting for obvious degeneracies or (non-linear) combinations that simplify the problem, e.g. $q_0$.

Finally, the (log-)ratio estimator combines the featurized parameter(s) with the output of the data head. Since global parameters are informed only by summary statistics, the respective network only considers $\mathcal{S}$, while for latent variables of a particular SN, the (featurized) observational data relating to that object is also passed as input:

$$\begin{aligned} \text{global:} \ \ln \hat{r}(\boldsymbol{\theta}_i, \boldsymbol{d}) &= \mathtt{RatioEstimator}_{\boldsymbol{\theta}_i}(\boldsymbol{\theta}_i, \mathcal{S}), \\ \text{latent:} \ \ln \hat{r}(\boldsymbol{\vartheta}_i^s, \boldsymbol{d}) &= \mathtt{RatioEstimator}_{\boldsymbol{\vartheta}_i^s}(\boldsymbol{\vartheta}_i^s, \mathcal{S}, \boldsymbol{d}^s). \end{aligned} \tag{13}$$

Notice that the latent-variable ratio estimator needs to be parametrized by $s$, i.e. it needs to know which object's parameters are being inferred. Below, we discuss how we implement this parametrization so that we can infer all latent parameters with a single network.

*3.2.3 Auxiliary input for latent-variable inference*

In our current model, the data $[\boldsymbol{d}^s]_{s=1}^N$ are not invariant under permutation of the supernovae because each has a different sampling distribution controlled by $\hat{\boldsymbol{\Sigma}}^s$ and $\hat{z}^s$. Thus, for marginal latent-variable ratio estimation, we either need to train a separate ratio estimator for each SN Ia or parametrize the $s$-dependence of the ratio with an auxiliary *label*, $\boldsymbol{a}^s$, to identify which object is being considered. In principle, $\boldsymbol{a}^s$ could be as simple as the index $s$ itself, but since we know what the ratio estimator is supposed to learn: the observational covariance and observed redshift of the supernova, we simplify its task by directly setting $\boldsymbol{a}^s$ to a concatenation of the components of $\hat{\boldsymbol{\Sigma}}^s$ and $\hat{z}^s$ into a $3 \times 3 + 1$-dimensional vector.[9] We note that $\hat{\boldsymbol{\Sigma}}$ and $\hat{z}$ are, in reality, derived from the observed light curves and spectra, just like $\hat{m}$, $\hat{x}_1$, and $\hat{c}$. In our simplified model, considering them as part of the data, even though they are fixed in the simulator, makes each supernova fully identifiable.

For added expressivity, we first featurize the labels with an auxiliary head network similar to the data pre-processor:

$$\boldsymbol{a}^s = \mathtt{AuxiliaryHead}(\boldsymbol{a}^s), \tag{14}$$

before appending them to the featurized data as input to the ratio estimator.[10]

Finally, we note that the marginal posterior of latent variables of a given object depends, in general, on the observed data on *all* objects because of the possibility of correlations introduced by the likelihood and/or a priori in the forward model. As mentioned, this is accounted for in the inference network by $\mathcal{S}$ being an input also to latent-variable ratio estimators.

In our particular model, however, the data $\boldsymbol{d}^s$ on each individual SN Ia are independent since we take $\hat{\boldsymbol{\Sigma}}$ to be block-diagonal (see Section 2.5), and so a posteriori correlations between latent variables arise only because of the hierarchical structure: conditioned on particular values for the global parameters, the latent variables are indeed a priori uncorrelated. Similarly, when in the last stages of truncation (described in Section 3.3) the priors of the global parameters are tightly constrained, the latent variables are approximately uncorrelated. In fact, truncating the prior ranges of global parameters can be viewed as an approximate way of accounting for the effect that measuring $N-1$ other SNe Ia has on inferring the parameters of each one SN Ia – which is precisely the purpose of the data set summary, $\mathcal{S}$. Hence, in our experiments, including the connections from $\mathcal{S}$ to the latent-variable ratio estimator makes little difference to the inferred posteriors. We, instead, present results where these connections have been expunged, which simplifies greatly the network and hastens its training.

*3.2.4 NN implementation*

We implement all components of the inference network with multilayer perceptrons (MLPs), each of which contains a number of layers, as detailed in Table 2. Each layer consists of a fully connected linear part whose output is whitened online, i.e. shifted by the mean and rescaled by the standard deviation of hitherto-seen examples.[11] The mean and standard deviation of each layer's outputs over the training data are, thus, learnt parameters updated at every training step (albeit not with gradient descent) and then fixed during inference. Finally, a rectified linear unit (ReLU) non-linearity is applied (except for the output layer). We found that the choice of non-linearity did not affect the results and chose the other network hyperparameters (number and size of layers) manually based on the final loss value and how fast the network learns in the initial stages of training.

Using only MLPs imposes restrictions on the simulator and the analysed data. In particular, the data on individual objects must be fixed in size and the same across all of them so that it can be fed to a fully connected linear layer, whereas the cadence of light curves typically varies from object to object. Similarly, using a fully connected summarizing component requires the size of the simulated sample to be exactly the same throughout training and, hence, that the simulator be conditioned on the number of observed SNe Ia, but selection effects and differing levels of contamination lead to a variable $N$ in the simulator. Another consequence is that the network size scales with the data set, reaching about a billion trainable parameters and a memory footprint of $\sim 10$ GB for $10^5$ SNe Ia.

---

[9]In fact, for convenience of implementation, we provide the Cholesky factor of $\hat{\boldsymbol{\Sigma}}^s$, which has three constant null – instead of duplicated – entries.

[10]Thus, in Table 2, we write the $s$-specific ratio estimator $\mathtt{RatioEstimator}_{\boldsymbol{\vartheta}_i^s}(\boldsymbol{\vartheta}_i^s, \mathcal{S}, \boldsymbol{d}^s)$ from equation (13) as $s$-agnostic but with an auxiliary input: $\mathtt{RatioEstimator}_{\boldsymbol{\vartheta}_i}(\boldsymbol{\vartheta}_i^s, \mathcal{S}, \boldsymbol{d}^s, \boldsymbol{a}^s)$.

[11]This is a special case of 'batch normalization' (Ioffe & Szegedy 2015) from the NN literature with momentum set so as to calculate cumulative averages in an online-learning setting.

**Table 2.** Details about the components of the inference network: their input and output dimensions and particular implementation in this work. For all components we use multilayer perceptrons (MLPs), for which we list the number and size of the hidden layers. Each hidden layer consists of a fully-connected layer, an online whitening step, and a rectified linear unit (ReLU) non-linearity. Inputs are also whitened online. The dimension of global parameter groups is denoted with $m = 1$, or 2 for $\theta_1 \equiv \mathcal{C}$. Note that in this work we did not provide the summary to the latent-variable ratio estimator, as motivated in Section 3.2.3. The inference network is also depicted in Fig. 2.

| Component | Inputs | $\in$ | Space | $\rightarrow$ | Output | $\in$ | Space | Implementation |
|---|---|---|---|---|---|---|---|---|
| DataHead | $d^s$ | $\in$ | $\mathbb{R}^3$ | $\rightarrow$ | $d^s$ | $\in$ | $\mathbb{R}^{32}$ | MLP($3 \times 128$) |
| Summariser | $[d^s]_{s=1}^N$ | $\in$ | $\mathbb{R}^{32 \times N}$ | $\rightarrow$ | $\mathcal{S}$ | $\in$ | $\mathbb{R}^{256}$ | MLP($2 \times 256$) |
| ParamHead$_{\theta_i}$ | $\theta_i$ | $\in$ | $\mathbb{R}^m$ | $\rightarrow$ | $\theta_i$ | $\in$ | $\mathbb{R}^{256}$ | MLP($2 \times 256$) |
| RatioEstimator$_{\theta_i}$ | $\theta_i, \mathcal{S}$ | $\in$ | $\mathbb{R}^{256+256}$ | $\rightarrow$ | $\ln \hat{r}(\theta_i, d)$ | $\in$ | $\mathbb{R}^1$ | MLP($3 \times 256$) |
| ParamHead$_{M_0}$ | $M_0^s$ | $\in$ | $\mathbb{R}^1$ | $\rightarrow$ | $M_0^s$ | $\in$ | $\mathbb{R}^1$ | Identity |
| AuxiliaryHead | $\hat{\Sigma}^s, \hat{z}^s$ | $\in$ | $\mathbb{R}^{9+1}$ | $\rightarrow$ | $a^s$ | $\in$ | $\mathbb{R}^{16}$ | MLP($3 \times 128$) |
| RatioEstimator$_{M_0}$ | $M_0^s, d^s, a^s$ | $\in$ | $\mathbb{R}^{1+32+16}$ | $\rightarrow$ | $\ln \hat{r}(M_0^s, d)$ | $\in$ | $\mathbb{R}^1$ | MLP($3 \times 256$) |

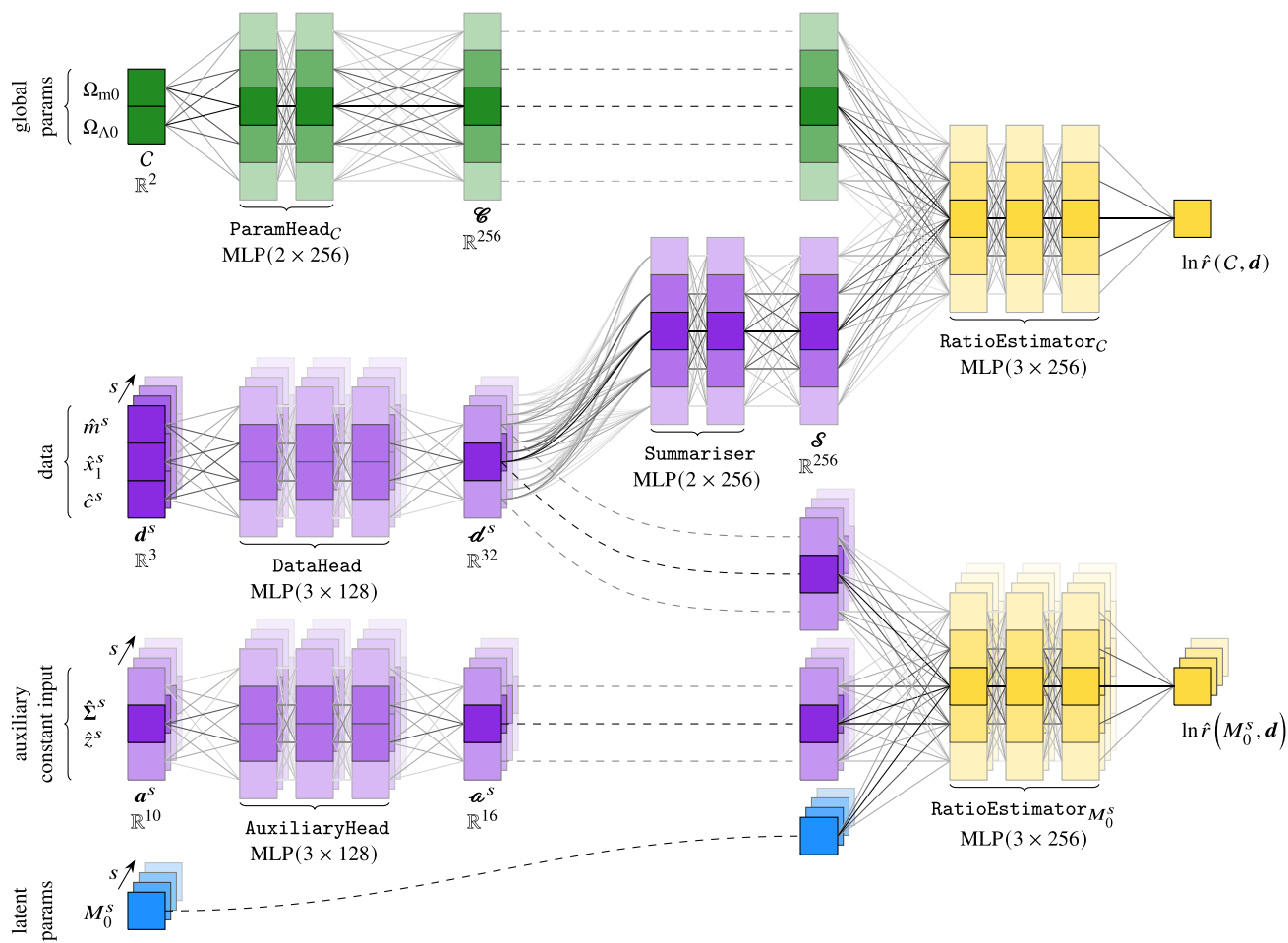

**Figure 2.** Inference network architecture as used for final inference, i.e. after constraining the non-cosmological global parameters. The solid lines represent linear connections, which inside the MLPs are followed by online whitening and a ReLU non-linearity. Inputs are also whitened. Dashed lines are an identity operation, duplicating a layer for presentation purposes. Note that in this work we expunge the connections from the summary to the latent-variable ratio estimator, as motivated in Section 3.2.3, and do not pre-process (featurize) the latent $M_0^s$. Full details on the specific components, inputs, outputs, and intermediate variables are given in Table 2.

We have explored alternatives to MLPs that do not suffer from these limitations; for example, 1D convolution along the SN Ia index dimension, recurrent and set-based architectures, whose application to realistic data will be presented in future work.

### 3.3 Truncated marginal ratio estimation

The key to practical ratio estimation is the ability of NNs to interpolate the function they are trained to approximate. Thus, their performance depends, on the one hand, on the amount of training they receive, i.e. on how densely the parameter space is sampled to produce training examples, and on the other, on the network's intrinsic capacity to represent the full complexity of the target. To alleviate the burden, Miller et al. (2020, 2022) proposed *truncated* marginal neural ratio estimation (TMNRE), which focuses the training on to regions of parameter space compatible with a specific *target* data set by successively (in stages) restricting the prior range from which training examples are generated. Unlike other sequential methods that use a posterior estimate as proposal for generating subsequent training data (Durkan, Papamakarios & Murray 2018; Durkan, Murray & Papamakarios 2020), this truncation scheme does not modify the shape of the prior distribution but merely restricts its support (and hence introduces to it only a uniform re-normalization).

At each stage, the truncated (excluded) region – determined independently for each parameter group – is that in which the posterior density is deemed negligible. Specifically, we restrict the prior to a rectangular box enclosing the highest probability density (HPD) region that contains 99.99 per cent (i.e. $1 - 10^{-4}$) probability mass from the current approximate posterior, $q(\Theta \,|\, d)$ (evaluated for the target data).[12] After each truncation, a new network is randomly initialized and trained on samples generated with the newly constrained priors.

### 3.3.1 Truncation and latent variables

In principle, the priors of latent variables can also be truncated iteratively, which would tailor the simulator output to the concrete properties of the observed SNe Ia. However, the *effective* priors of variables in intermediate layers of a hierarchical model (resulting from marginalizing the upstream model) are typically intractable or hard to sample from directly – this is precisely why the forward simulator is implemented in layers. Hence, sampling from a constrained latent-variable prior usually requires a rejection strategy, whose wastefulness increases the more the prior is truncated.

In this work, we do not explicitly restrict the prior ranges of latent variables. Instead, we only train a ratio estimator for them after we have constrained the global parameters as much as possible, i.e. after the last truncation stage. This results in simulations which resemble the targeted data set only in terms of the *distribution* of latent variables rather than particular values. Naturally, this modifies the range *and shape* of the effective latent-variable prior. However, the modifications are restricted to regions where the *joint* posterior is negligible (and hence so are all the marginals): restricting the global parameters can be viewed as a way to enact constraints in high dimensions.

## 3.4 Validation and calibration of amortized inference

Neural ratio estimation is an *amortized* technique: after the upfront cost of training the ratio estimator, inference can be performed quickly with any number of data sets. Put otherwise, in contrast to likelihood-based methods that learn the posterior *for a given* data set, ratio estimation learns the *procedure* to derive a posterior, and this procedure can be validated (in a Bayesian sense) with collections of simulated data and/or calibrated so as to produce (frequentist) confidence regions with guaranteed exact coverage, as we explain below.

---

[12] In one dimension, the box, i.e. a line segment, fits the HPD region exactly (except for multimodal posteriors, which we do not encounter in this work), but in higher dimensions, e.g. for the cosmological parameters $\mathcal{C}$, which we treat as one 2-dimensional parameter, this means the new constrained region usually contains even more probability mass than required.

In contrast to other sequential methods, inference using the truncation scheme described above is still *locally* amortized since the target data (and the posterior associated with it) is used during training only to determine the sequence of constraints in parameter space, while training data is always sampled according to the prior. Therefore, the inference network used in each stage of truncation can, and must, be validated and/or calibrated within the constrained region over which it has been trained.

### 3.4.1 Bayesian and frequentist coverage (P–P plots[13])

We wish to examine the empirical (i.e. determined from analyses of simulated data $d$) coverage properties of an inference procedure that derives approximate posteriors $q(\Theta \,|\, d)$. To do this, we first associate with each parameter value $\Theta_0$ a credibility $\gamma(\Theta_0, d)$ as the approximate posterior probability enclosed by the highest probability density (HPD) region which has $\Theta_0$ on its boundary:

$$\gamma(\Theta_0, d) \equiv \int_{\Gamma_\Theta(\Theta_0, d)} q(\Theta \,|\, d) \, d\Theta, \tag{15}$$

where the integration is over the region where the approximate posterior density is higher than at $\Theta_0$:

$$\Gamma_\Theta(\Theta_0, d) \equiv \{\Theta : q(\Theta \,|\, d) > q(\Theta_0 \,|\, d)\}. \tag{16}$$

See also the top panel of Fig. 3 for an illustration. Taking $\Theta_0$ to be the true parameters used to generate $d$, we then plot the frequency $F(\gamma)$ with which HPD regions of different credibility $\gamma$ include (cover) $\Theta_0$. If this empirical coverage is larger than the credibility ($F(\gamma) > \gamma$), the approximation is said to be conservative: it covers more frequently than its credibility; if $F(\gamma) < \gamma$, on the other hand, it is said to be undercovering.

In practice, such a plot is built by repeatedly simulating data $d$ from parameters $\Theta_0$, deriving $q(\Theta \,|\, d)$ with the inference procedure being validated, determining from it the region $\Gamma_\Theta(\Theta_0, d)$ where $q(\Theta \,|\, d) > q(\Theta_0 \,|\, d)$, and integrating $q(\Theta \,|\, d)$ over it, as illustrated in the top panel of Fig. 3. The cumulative distribution of the $\gamma(\Theta_0, d)$ obtained in this way gives the empirical coverage $F(\gamma)$. Full details are presented in Appendix A.

There are two possible options for choosing the distribution of parameters[14] from which to simulate data in the first place, corresponding to Bayesian and frequentist approaches.

Bayesian validation through the empirical distribution of credibilities was discussed by Cook, Gelman & Rubin (2006), Talts et al. (2020), and more recently, in the context of LFI and NRE, by Hermans et al. (2022). In this setting, the so-called self-consistency of the prior and data-averaged posterior (see Appendix A1) means that the empirical coverage frequency of a set of credible regions matches their credibility calculated with the exact posterior. Therefore, by examining deviations from the diagonal line in a Bayesian P–P plot, one can determine whether the NRE posteriors $q(\Theta \,|\, d)$ are good approximations to the exact posterior $p(\Theta \,|\, d)$ and, if not, whether the inference procedure is conservative or undercovering, *on average* across the prior range. Still, $F_B(\gamma) = \gamma$ (i.e. a diagonal P–P plot) is only a necessary but not sufficient condition for $q(\Theta \,|\, d) \to p(\Theta \,|\, d)$; indeed, even $q(\Theta \,|\, d) = p(\Theta)$ has proper Bayesian coverage (i.e. the prior is self-consistent with the data-averaged *prior*).

---

[13] See e.g. Gibbons & Chakraborti (2010, section 4.8).
[14] This concerns the parameters of interest. The rest are always drawn from their (conditional) priors.

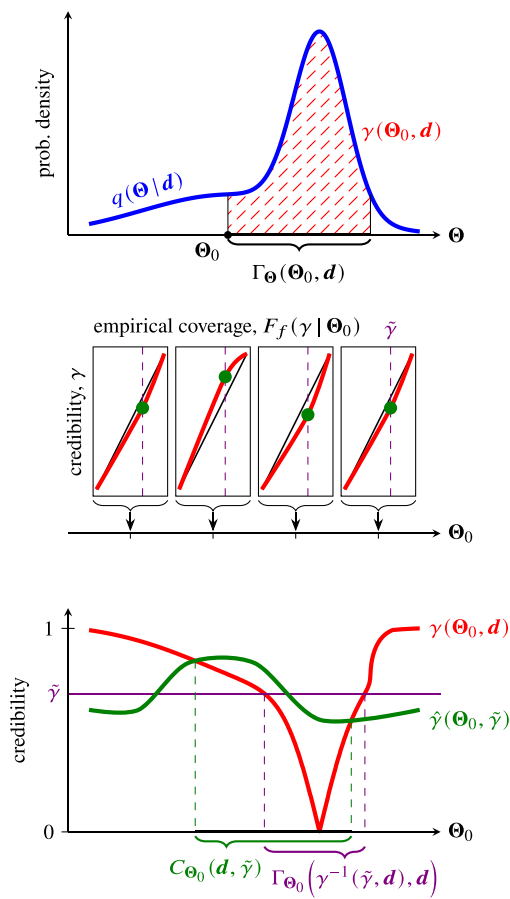

**Figure 3.** Procedure for obtaining frequentist confidence regions with exact coverage from an approximate amortized Bayesian posterior. Top panel: the credibility $\gamma(\Theta_0, d)$ associated with a parameter value $\Theta_0$ is the credibility of the HPD region bounded by $\Theta_0$ for a given (approximate) posterior $q$, conditioned on data $d$. Middle panel: by repeated draws of $d$ at fixed $\Theta_0$, we obtain samples for $\gamma(\Theta_0, d)$, from which we build its empirical cumulative distribution, $F_f(\gamma | \Theta_0)$ (red lines). The credibilities $\hat{\gamma}(\Theta_0, \tilde{\gamma})$ required for regions to cover $\Theta_0$ with a given frequency $\tilde{\gamma}$ are indicated with green dots and determined as the $\tilde{\gamma}$th quantiles of $F_f$. We obtain the green line in the bottom panel by repeating this procedure on a grid of $\Theta_0$. Bottom panel: $C_{\Theta_0}(d, \tilde{\gamma})$, the region with confidence level $\tilde{\gamma}$, is that in which $\gamma(\Theta_0, d)$ (red line) is lower than $\hat{\gamma}(\Theta_0, \tilde{\gamma})$ (green line). For comparison, if $q$ is equal to the true posterior, and a uniform prior is used, $\hat{\gamma}(\Theta_0, \tilde{\gamma})$ is constant across all $\Theta_0$ and equal to the target confidence level (purple line). The credible region, $\Gamma_{\Theta_0}(\gamma^{-1}(\tilde{\gamma}, d), d)$, then coincides with the confidence region.

On the other hand, if data are simulated at fixed parameter values, i.e. from $p(d | \Theta)$, one is working in a frequentist setting. If the prior is uniform, credible regions of the marginal Bayesian posterior again cover with frequency equal to their credibility, but this property is not guaranteed a priori.

In any case, irrespective of the prior used and even from an approximation to the posterior, one can still derive exact confidence regions, as demonstrated by Dalmasso, Izbicki & Lee (2020), Dalmasso et al. (2022), and Masserano et al. (2022). While traditionally the focus has been on likelihood ratios as a test statistic, motivated by the Neyman–Pearson lemma, we consider $\gamma(\Theta_0, d)$ instead. If the prior in $\Theta$ is uniform, the two choices are equivalent since $\gamma(\Theta_0, d)$ is monotonic in the posterior probability density and thence, in the likelihood. Otherwise, our procedure takes into account the effect that the prior has on posterior credibility and *corrects* for it implicitly.

Building confidence sets from credibilities is then similar to the usual Neyman construction (Neyman & Jeffreys 1937): first, a series of 'frequentist' validation plots for different fixed parameters across the support of the prior are built; from them, one can derive a map of 'threshold credibilities', $\hat{\gamma}(\Theta_0, \tilde{\gamma})$, for any given confidence level $\tilde{\gamma}$; finally, for a particular data set $d$, the confidence region includes parameter values that attain in $q(\Theta | d)$ a *lower* credibility (i.e. a less extreme test statistic) than the threshold. The procedure is illustrated in the middle and bottom panels of Fig. 3 and fully detailed in Appendix A2.

## 4 EXPERIMENTS AND RESULTS

In this section, we apply our inference procedure to mock data simulated by the forward model described in Section 2 and depicted in Fig. 1. Even though the likelihood in this case is known, obtaining posteriors for the cosmological parameters of interest with traditional methods would in general require explicitly sampling all individual SN Ia parameters, which is impractical for the $\sim 10^5$ SNe Ia expected from future surveys. Instead, one might attempt to analytically marginalize the latent variables in order to reduce the parameter space to just the $\mathcal{O}(10)$ global parameters. This is possible for our current model, in which the latent-variable priors and the likelihood are normal, with the only necessary approximation being a linear propagation of the redshift uncertainties to the distance modulus, as detailed in Appendix B. However, this simplification can significantly bias inference of the cosmological parameters (see Fig. 4). The bias, which increases in severity for larger sample sizes, arises due to abundance in the sample of supernovae at low redshift, where the non-linearity of the distance modulus is most pronounced: using an incorrect likelihood then favours a different shape of $\mu(\mathcal{C}, z)$ and, hence, different cosmological parameters. Furthermore, propagating uncertainties only linearly artificially enlarges the expected variability, which, on the other hand, leads to overconfident results (evidenced by the tightly constraining, yet biased, MCMC posteriors in Fig. 4) when these variations are not present in the data. In contrast, marginal NRE places no restrictions on the model and, therefore, produces unbiased posteriors with correct uncertainties, as shown in Fig. 4.

In the following, we first elaborate on the application of TMNRE to infer global model parameters, validate the approximate posteriors, and produce calibrated frequentist confidence regions in the space of cosmological parameters. We then validate the NRE posteriors against MCMC results for mock data generated with the simplified tractable model, which, however, is a poor description of *real* data. Finally, we demonstrate high-dimensional NRE on the latent parameters, specifically, on the intrinsic magnitudes $M_0^s$.

### 4.1 Inferring global parameters with iterative truncation

We apply TMNRE to mock data sets containing between $10^3$ and $10^5$ SNe Ia, all sharing the same global parameter values listed in Table 1 but with different latent parameters. For each sample size we also generate a collection of observed redshifts, $\hat{z}$, and observational covariance $\hat{\Sigma}$ as described in Section 2.5. For each simulator configuration (uniquely defined by $\{N, \hat{z}, \hat{\Sigma}\}$), we train separate inference networks.[15] The analysis proceeds in *stages*, starting with the priors listed in Table 1 and truncating them as described in Section 3.3, targeting the mock observation. We stop truncating when none of the parameter ranges shrink by more than a factor of 2.

---
[15]In the future, we plan to modify the network structure so as to accept arbitrary-size permutation-invariant data sets, which will make the inference network completely general.

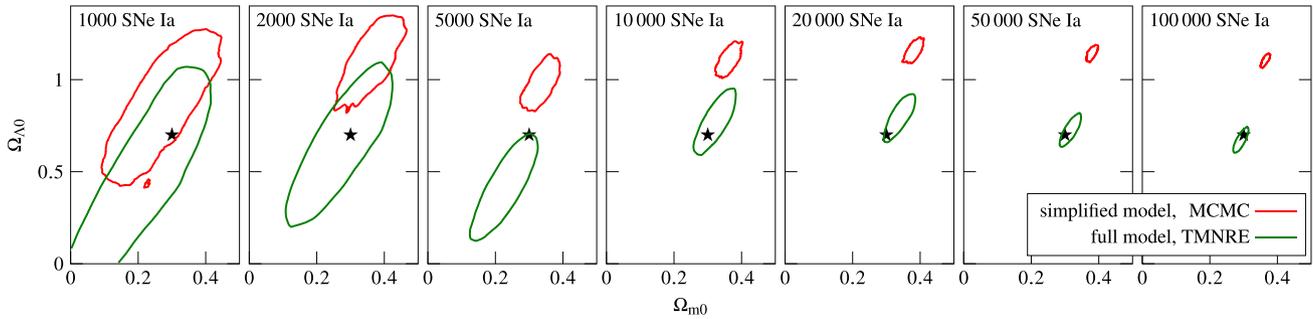

**Figure 4.** Marginal posteriors for cosmological parameters (90 per cent credible regions) for increasing SN Ia sample size from TMNRE (with the full generative model) in green and from MCMC (with a linearized model, required to make the problem tractable for sampling methods) in red. A star marks the values used to simulate mock data.

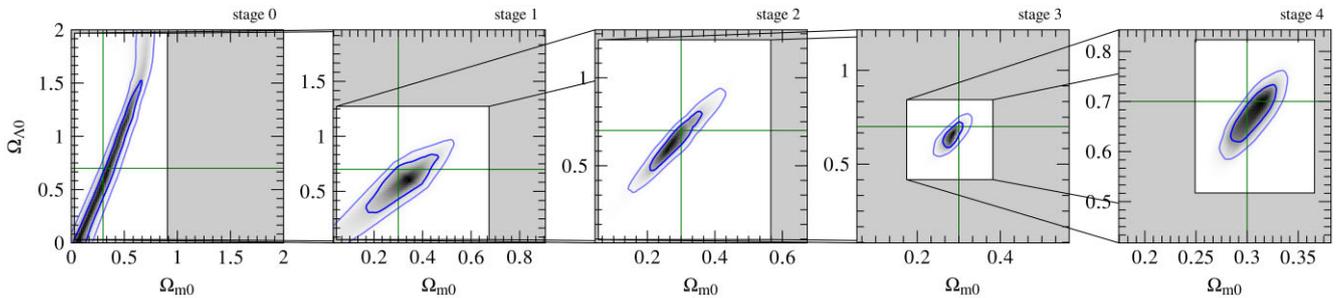

**Figure 5.** Approximate posterior density and 68 per cent and 95 per cent HPD contours for the cosmological parameters of interest at sequential stages of truncation in the analysis of $10^5$ SNe Ia. The rectangular boxes delineate the selected range for the following stage. The prior used is always flat across the depicted range, and green lines denote the values used to produce the mock data.

For training we use the Adam optimizer (Kingma & Ba 2017) with initial learning rate of $10^{-4}$, decreasing it by a factor of 2 every 2000 steps for a total of 10 000 steps. We do not use a fixed training set but rather produce new examples on-the-fly in mini-batches of size 64, so that the total number of simulations used to train a given network (i.e. the estimator *for a given stage of truncation*) is 640 000. New training examples are generated at each stage. Training in the case of $10^5$ SNe Ia took $\approx$2 h per stage (around 12 h in total) on an NVIDIA A-100 GPU. The run time is dominated by the simulator, whereas the memory footprint is largely due to the network, specifically, the dense summarizing layer.

The results are illustrated in Figs 5 and 6 for the data set with $10^5$ SNe Ia. Often, at first not all parameters are constrained: e.g. $\alpha$ and $\beta$ are learnt appreciably only after two truncations, when the training data variance is reduced enough by constraining the rest of the parameters. On the other hand, the obvious approximate degeneracy in the cosmological parameters ($q_0 = \Omega_{m0}/2 - \Omega_{\Lambda0}$) is immediately picked up by the network, as indicated by the narrow strip in the first panel of Fig. 5. Note that the posteriors in the initial stages are much larger than later ones even though training has largely converged (as evidenced by the loss stabilizing): this is due to the limited network capacity and demonstrates the need and utility of truncation.

The final cosmological posterior we obtain by training with the final truncated priors for all global parameters but only inferring the cosmological parameters in order to utilize the full flexibility of the data pre-processor network.

This same procedure was used for all the NRE posteriors in Fig. 4.

### 4.2 Validation and calibration

We validate the posteriors from the last truncation stage in the analysis of $10^5$ SNe Ia, i.e. those depicted in the last panels of Figs 5 and 6, using a Bayesian P–P plot shown in Fig. 7. We use 4096 mock data sets, which the trained NN analysed in $\approx$1 min. For all parameter groups the empirical coverage is close to the credibility of the posterior approximation, as indicated by the nearly diagonal lines. The biggest deviations, of up to 5 per cent, are in the cosmological parameters, which is to be expected since it is the only 2D posterior we are inferring. This provides further motivation for including an additional training phase targeting exclusively cosmology.

We also validate the final cosmological posterior approximation at fixed parameter values. The P–P plots from a grid across the final truncated prior (the box in the last panel of Fig. 5) are shown in Fig. A1, and the full map of the credibility required to cover with 68.4 per cent confidence is shown in Fig. 8b. Using this map and the procedure described in Section 3.4.1 and Appendix A2 we derive and show in Fig. 8a confidence regions with exact coverage from the approximate posteriors for three different data sets generated with parameters randomly drawn from the (truncated) priors. We notice that, even though the approximation is in some regions conservative and in others undercovering by a few tens per cent (for 68 per cent coverage), this is actually the result of small inaccuracies in the size of the inferred posterior, which would not affect scientific conclusions and can in any case be calibrated away.

### 4.3 Validation of NRE posteriors on a tractable example

In this subsection, we test the NRE inference procedure against 'ground-truth' posteriors obtained with MCMC. In order to make the latter numerically tractable, we marginalize out the latent layer of BAHAMAS and linearize redshift error propagation, as described in Appendix B. From this simplified model, we generate a number of data sets, varying the number of SNe Ia between $N = 10^3$–$10^5$, spanning the range from current to near-future surveys. We mimic spectroscopic observations by setting $\sigma_z = 0$ (labelled specz) in addition to data sets with photometric redshift errors propagated linearly ($\sigma_z = 0.04$, a variable to be inferred, labelled mphotoz). In each case, we generate 10 data realizations with the same global parameters (from Table 1) but different latent parameters.

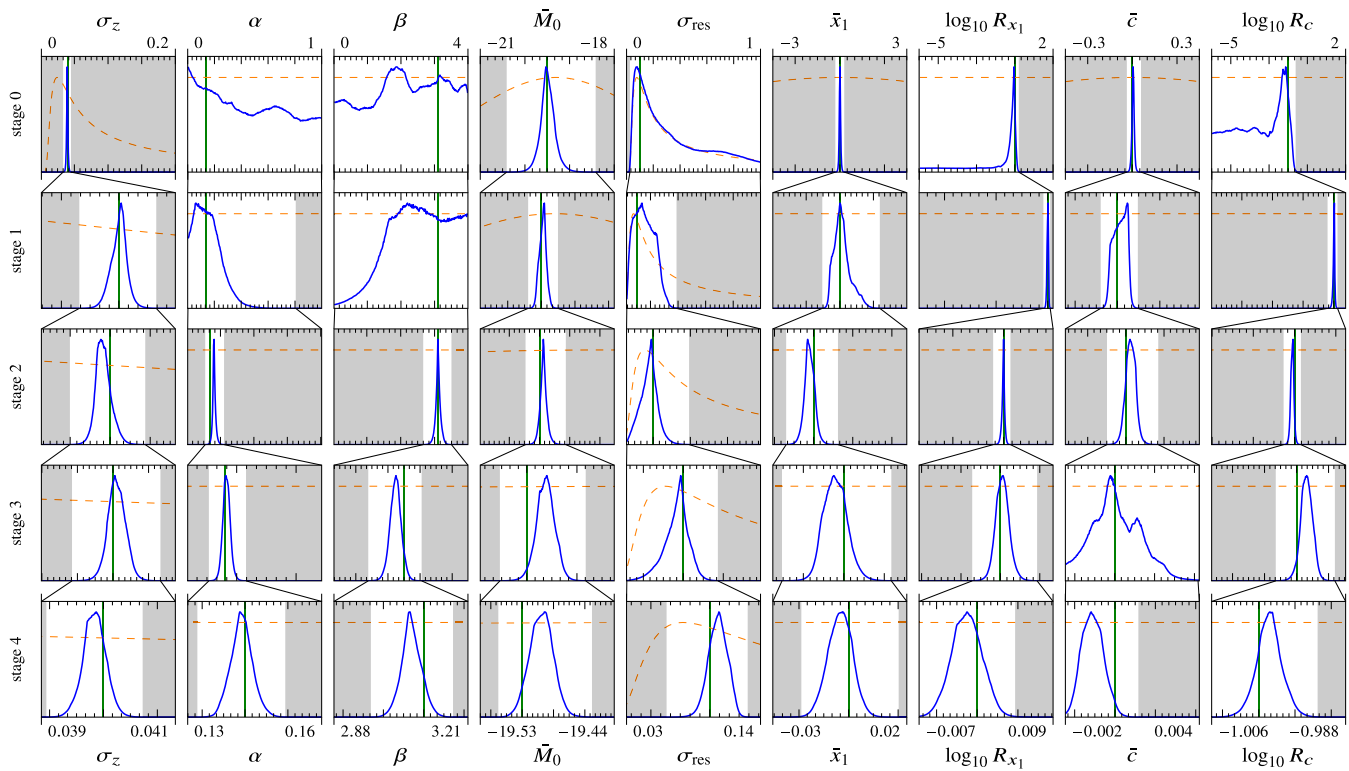

**Figure 6.** Approximate posteriors (blue solid lines) for the non-cosmological global parameters at sequential stages of truncation (each row is a stage) in the analysis of $10^5$ SNe Ia. The dashed lines show the prior density (the same across stages), which gets truncated to the unshaded region for the following stage. In all plots the green vertical line denotes the value used to produce the target mock observations (from Table 1), and the approximate posteriors and priors are independently arbitrarily normalized to aid presentation.

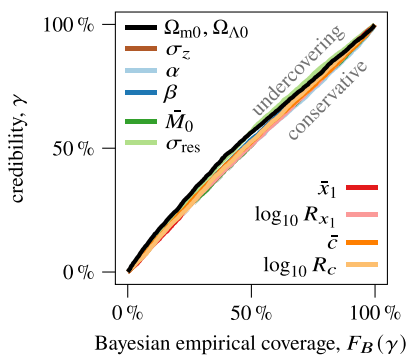

**Figure 7.** Bayesian P–P plots for the 10 marginal posteriors inferred from $10^5$ SNe Ia in the last stage of truncation. (The posteriors themselves evaluated on the target mock observations are shown in Figs 5 and 6.) Note that the axes are switched with respect to other works (e.g. Hermans et al. 2022).

We perform the MCMC analysis using the EMCEE package (Foreman-Mackey et al. 2013), sampling all 11 global parameters (10 for specz) for 1000 steps (discarding the first 200 as burn-in) with 50 chains.

For each sample size $N$ and data type (specz or mphotoz) we then train a neural ratio estimator for the cosmological parameters only. We imitate the last stage of iterative truncation by constraining the prior of each global parameter so that it contains (to within $5\sigma$) the MCMC posteriors from all 10 analysed data sets. Once the inference network is trained, the 10 NRE posteriors can easily be evaluated.

We present a selection of posteriors from $10^5$ SNe Ia in Fig. 9. We observe that NRE posteriors are slightly larger than and slightly displaced from the MCMC 'ground truth', although there is no systematic bias. The displacement appears to be larger when the true parameters fall in the tail of the true posterior, i.e. for more 'unlikely' data, of which the inference network has seen fewer examples during training. Still, examining the Bayesian P–P plots, which are very similar to Fig. 7, reveals a slight undercoverage, which means that even though the approximate posteriors are slightly larger, they are scattered around the true values. We also remind the reader that Bayesian P–P plots indicate performance across the whole parameter space, while in Fig. 9 we only show posteriors on data generated from a single set of parameter values.

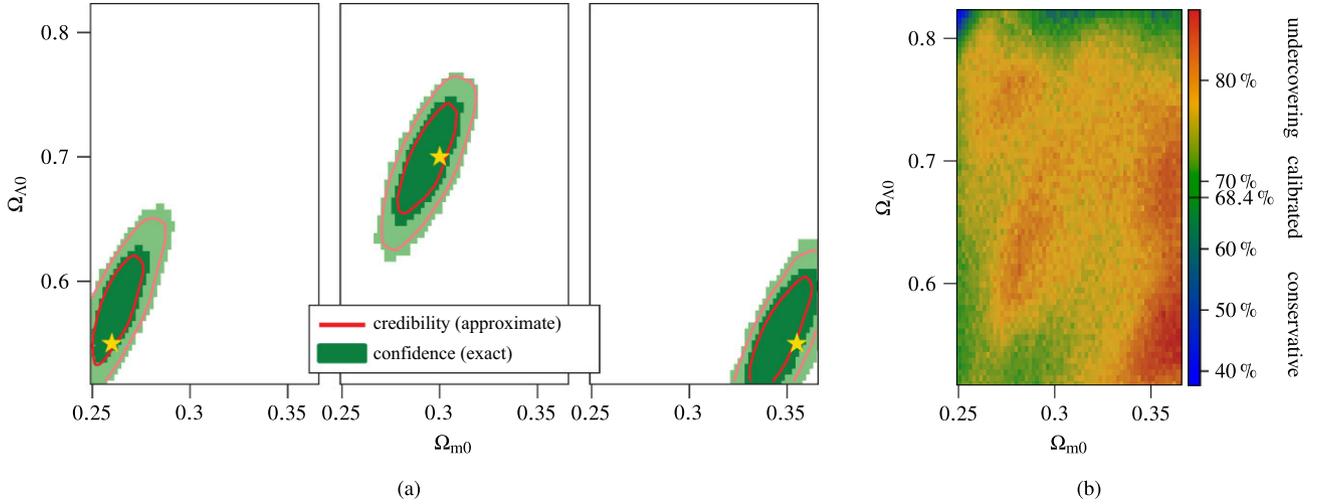

**Figure 8.** (a) Calibrated constraints from $10^5$ SNe Ia with different input cosmological parameters (indicated by a star). Red contours delineate the 68 per cent and 95 per cent (approximate) credible regions from the last stage of TMNRE (whose constrained prior range is the extent of the plots), while the respective calibrated (exact) confidence regions are shown in green. (b) Credibility of the approximate posterior required to cover the true value for 68.4 per cent of data realizations (in a frequentist sense, i.e. at fixed values of the cosmological parameters according to the location of the pixel).

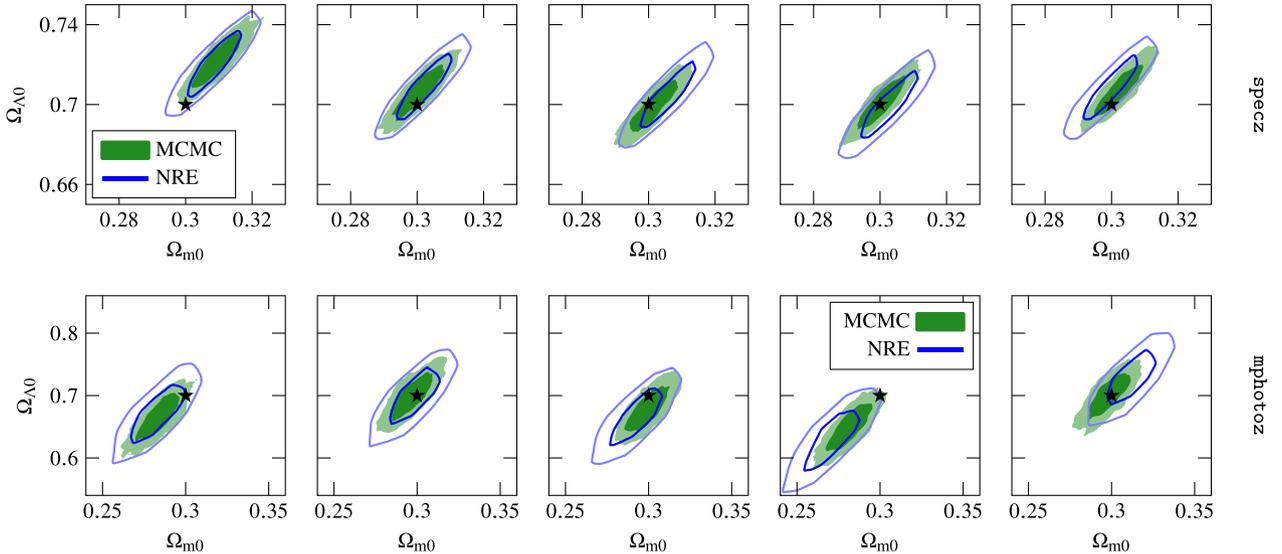

**Figure 9.** Comparison of posteriors for the cosmological parameters (68 per cent and 95 per cent HPD credible regions) from NRE (blue contours) and MCMC (green filled areas) for $10^5$ SNe Ia simulated with the linearized marginal BAHAMAS model (see Appendix B), for which the MCMC solution is taken as the ground truth. On the top row are five different mock data realizations with spectroscopic redshifts ($\sigma_z = 0$), and on the bottom with $\sigma_z = 0.04$. Note the different scales for the two rows. A star shows the true parameters used to generate the mock data.

Finally, in Fig. 10, we examine how the precision and accuracy[16] of NRE posteriors behave across the range of SN Ia sample sizes. We confirm that NRE posteriors are consistently larger than their MCMC counterparts by up to a few tens per cent per parameter. Still, both MCMC and NRE posterior sizes clearly scale as $1/\sqrt{N}$ per parameter, while the offset of the mean is proportional to the standard deviation for both MCMC and NRE analyses, as expected in a purely Gaussian model. This means that the inference network succeeds in combining the information from the large number of observed objects. Overall, NRE achieves comparable accuracy and precision to MCMC across the sample sizes considered, and we do not observe signs of bias.

### 4.4 Inference for individual SNe Ia

To infer latent parameters with NRE, one simply needs to designate them as parameters of interest and feed them in the ratio estimator, as described in Sections 3.2.2 and 3.2.3. However, since the effective prior of latent variables is only defined implicitly through the hierarchical model, it is intractable, and so the posterior density cannot be directly evaluated numerically. Instead, we use simulator samples (the ones used in training) to represent the prior and re-weight them using the neural ratio estimate. We can then calculate moments or build histograms to represent the marginal NRE posteriors.

---

[16] We measure precision through the determinant of the approximate-posterior covariance: size $\equiv \sqrt{|\langle \mathcal{CC}^\mathrm{T} \rangle|}$, and accuracy with the offset of the mean from the ground-truth: bias$^2 \equiv |\langle \mathcal{C} \rangle - \mathcal{C}_\mathrm{true}|^2$, where averages are over $q(\mathcal{C} \mid \boldsymbol{d})$. Even though these two quantities have the same dimensions for our 2D cosmological posteriors, we caution against comparing them directly since the former represents a volume in parameter space, while the latter is the square of a length.

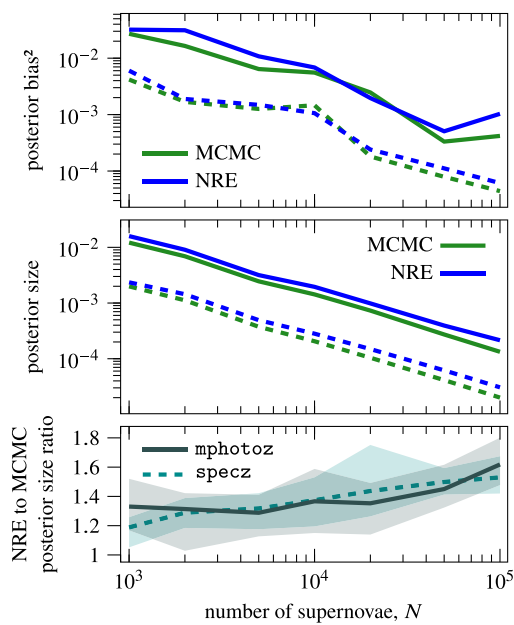

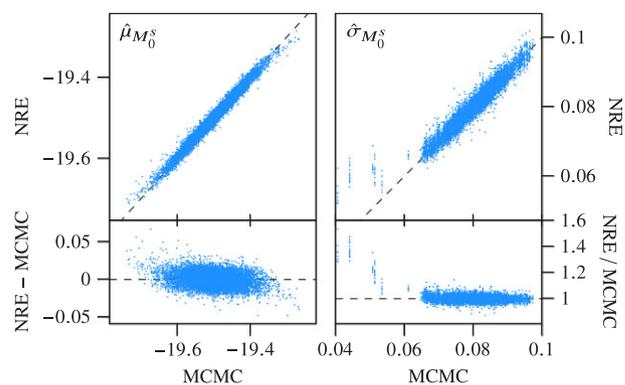

**Figure 10.** Comparison of the accuracy (top) and precision (middle), as defined in footnote 16, of NRE (blue) and MCMC (green) posteriors for data sets of different sizes generated and analysed with the linearized marginal BAHAMAS model (see Appendix B). The units in the top two panels are $[\mathcal{C}]^2$, and the general trend is $\propto 1/N$ (i.e. $1/\sqrt{N}$ *per dimension* in the 2D space of cosmological parameters), as expected from Gaussian theory. The bottom panel shows the ratio of posterior sizes from the two methods (i.e. of the respective lines in the middle panel). Everywhere solid lines pertain to mphotoz data and dashed lines to specz. All plots have been averaged over 10 data realizations, and in the bottom panel, the shaded areas additionally show the range of values across the different mock data sets.

We do not iteratively constrain the latent parameters and instead only infer them after the priors of the global parameters have been truncated. As discussed in Section 3.2.3, this allows us to simplify the inference network and not feed data set summaries into the latent-variable ratio estimator.

We present marginal results for the intrinsic magnitudes, $M_0^s$, of $N = 10^5$ SNe Ia simulated with spectroscopic redshifts (i.e. setting $\sigma_z = 0$) since data with photometric redshift uncertainty is too weakly constraining in the parameters of an individual SN Ia: the marginal posteriors nearly coincide with the constrained effective prior. In this case, as described in Appendices B and B1, we can also perform the global inference with MCMC and then use the samples to obtain latent-variable posteriors, which we consider as ground truth. In Fig. 11, we compare their moments[17] to moments of the marginal NRE posteriors.

NRE learns to correctly identify the location and size of the latent-variable posteriors. There are small deviations from the ground truth for extreme values: where the posterior (and thence, the data) falls in the tail of the prior, and so the network has seen few similar examples during training, and when the posterior is tightly constraining (i.e.

---

[17]The true posteriors in this case are Gaussian, and we have verified visually that the NRE posteriors resemble Gaussians.

**Figure 11.** Comparison of marginal posterior moments (left-hand panel: mean, and right-hand panel: standard deviation) for the intrinsic brightnesses of SN Ia, $M_0^s$, derived with NRE and MCMC. The values themselves are plotted in the top row, while the bottom panels show the difference of means and ratio of standard deviations. The data set analysed contains $10^5$ SNe Ia with spectroscopic redshifts as described in the text. All $10^5$ NRE posteriors were produced by one single inference network, whereas the MCMC results required an initial global sampling step and lengthy post-processing. The clustering of posterior variances at specific values is due to the sampling with replacement of the observational covariances when generating mock data (see Section 2.5).

has a small standard deviation). In the latter case, the NRE estimate is again conservative.

## 5 DISCUSSION AND CONCLUSIONS

We have presented a proof of concept for Supernova Ia Cosmology with truncated marginal neural Ratio EsTimation (SICRET). The main advantages of this modern likelihood-free method are that it scales to very large data sets (in this paper we analysed up to $10^5$ simulated SNe Ia), and that it allows seamless inclusion of arbitrary physical, statistical, and observational effects in the analysis. By reformulating Bayesian inference as a classification task, TMNRE allows a neural network trained on simulated data to implicitly take these effects into account and directly produce results for the limited set of parameters of interest: in our case, the matter and dark-energy densities. Meanwhile, an iterative truncation procedure tailors the simulator so that training data resembles the analysed (target) data set, which improves the network's learning.

After demonstrating the systematic bias that can be introduced by oversimplifying aspects of the data-generation process, e.g. uncertainties in measured redshift, we have presented posteriors for cosmological parameters derived with TMNRE from $10^5$ mock SNe Ia simulated from the BAHAMAS model. Exploiting the local amortization of TMNRE inference, we have validated the approximate posteriors for appropriate Bayesian coverage and presented a procedure to derive confidence regions with exact frequentist coverage. Furthermore, we have verified the TMNRE posteriors directly by comparing them with the 'ground-truth' obtained with standard MCMC on a simplified model (with spectroscopic redshifts or redshift uncertainty propagated linearly). We have shown that the inference network is able to extract all the relevant information even from large numbers of SNe Ia.

Moreover, we have presented a method to simultaneously infer the latent parameters of all $10^5$ SNe Ia with a single small neural network that takes advantage of the truncation of global-parameter prior ranges. We have verified the results for the intrinsic magnitude parameter in the simplified case against MCMC.

The likelihood-free simulator-based framework we have described will allow us to improve the simulator in the future, so that it is suitable for real data, without modifying the inference procedure, and scale it to the output of upcoming surveys like LSST. This will include modelling full SN Ia light curves using an underlying spectral energy distribution and a realistic instrumental model (e.g. calibration), accounting for selection effects, the influence of the environment through dust extinction, and the possibility of correlations with host properties, considering all contributions to the observed redshift and its uncertainty and scatter, and finally, non-SN Ia contaminants in the sample.

Future developments of the technique presented here might also help in shedding light on the mysterious tension in the value of the Hubble–Lemaître constant obtained from high- versus low-redshift probes.

## ACKNOWLEDGEMENTS

We would like to thank David van Dyk, Davide Piras, and Sebastian Goldt for useful comments and discussions and the anonymous referee for a careful read and insightful comments. The following software were used in this research: CLIPPPY,[18] a probabilistic framework based on PYRO (Bingham et al. 2019), for the generative model; PHYTORCH[19] for cosmological distance calculations (with gradients); and PYTORCH (Paszke et al. 2019) and PYTORCH LIGHTNING[20] for defining and training the neural networks.

## DATA AVAILABILITY

The mock data underlying this article will be shared on reasonable request to the corresponding author.

## REFERENCES

Abbott T. M. C. et al., 2019, ApJ, 872, L30
Betoule M. et al., 2014, A&A, 568, A22
Bingham E. et al., 2019, J. Mach. Learn. Res., 20, 973
Boone K., 2019, AJ, 158, 257
Boruah S. S., Hudson M. J., Lavaux G., 2020, MNRAS, 498, 2703
Boruah S. S., Hudson M. J., Lavaux G., 2021, MNRAS, 507, 2697
Brout D., Scolnic D., 2021, ApJ, 909, 26
Brout D. et al., 2019, ApJ, 874, 106
Brout D. et al., 2022, ApJ, 938, 110
Burke D. L. et al., 2018, AJ, 155, 41
Calcino J., Davis T., 2017, J. Cosmol. Astropart. Phys., 2017, 038
Childress M. et al., 2013, ApJ, 770, 108
Conley A. et al., 2011, ApJS, 192, 1
Cook S. R., Gelman A., Rubin D. B., 2006, J. Comput. Graph. Stat., 15, 675
Cranmer K., Brehmer J., Louppe G., 2020, Proc. Natl. Acad. Sci., 117, 30055
Dalmasso N., Izbicki R., Lee A., 2020, Proc. 37th Int. Conf. Mach. Learn., Confidence Sets and Hypothesis Testing in a Likelihood-Free Inference Setting. PMLR, Cambridge, MA, p. 2323
Dalmasso N., Masserano L., Zhao D., Izbicki R., Lee A. B., 2022, preprint (arXiv:2107.03920)
Davis T. M. et al., 2011, ApJ, 741, 67
Devroye L., Györfi L., Lugosi G., 1996, A Probabilistic Theory of Pattern Recognition, corrected edition. Springer, New York

Di Valentino E. et al., 2021, Class. Quantum Gravity, 38, 153001
Durkan C., Papamakarios G., Murray I., 2018, preprint (arXiv:1811.08723)
Durkan C., Murray I., Papamakarios G., 2020, Proc. 37th Int. Conf. Mach. Learn., On Contrastive Learning for Likelihood-Free Inference. PMLR, Cambridge, MA, p. 2771
Foley R. J., 2012, ApJ, 748, 127
Foley R. J. et al., 2018, MNRAS, 475, 193
Foreman-Mackey D., Hogg D. W., Lang D., Goodman J., 2013, PASP, 125, 306
Gardner J. R., Pleiss G., Bindel D., Weinberger K. Q., Wilson A. G., 2021, preprint (arXiv:1809.11165)
Gibbons J. D., Chakraborti S., 2010, Nonparametric Statistical Inference, 5th edn. Chapman and Hall/CRC, Boca Raton
Guy J., Astier P., Nobili S., Regnault N., Pain R., 2005, A&A, 443, 781
Guy J. et al., 2007, A&A, 466, 11
Guy J. et al., 2010, A&A, 523, A7
Hermans J., Begy V., Louppe G., 2020, Proc. 37th Int. Conf. Mach. Learn., ICML'20, Likelihood-free MCMC with amortized approximate ratio estimators. JMLR, Lille, France, p. 4239
Hermans J., Delaunoy A., Rozet F., Wehenkel A., Begy V., Louppe G., 2022, Trans. Mach. Learn. Res.
Hicken M. et al., 2009, ApJ, 700, 331
Hicken M. et al., 2012, ApJS, 200, 12
Hill R. et al., 2018, MNRAS, 481, 2766
Hinton S. R. et al., 2019, ApJ, 876, 15
Hlozek R. et al., 2012, ApJ, 752, 79
Huterer D., 2020, ApJ, 904, L28
Ioffe S., Szegedy C., 2015, Proc. 32nd Int. Conf. Int. Conf. Mach. Learn. Vol. 37, Batch Normalization: Accelerating Deep Network Training by Reducing Internal Covariate Shift. JMLR, Lille, France, p. 448
Ivezić Ž. et al., 2019, ApJ, 873, 111
Jennings E., Wolf R., Sako M., 2016, preprint (arXiv:1611.03087)
Jha S. et al., 2006, AJ, 131, 527
Jones D. O. et al., 2017, ApJ, 843, 6
Jones D. O. et al., 2022, ApJ, 933, 172
Kenworthy W. D. et al., 2021, ApJ, 923, 265
Kessler R., Scolnic D., 2017, ApJ, 836, 56
Kessler R. et al., 2009a, PASP, 121, 1028
Kessler R. et al., 2009b, ApJS, 185, 32
Kingma D. P., Ba J., 2017, preprint (arXiv:1412.6980)
Krisciunas K. et al., 2017, AJ, 154, 211
Kunz M., Bassett B. A., Hlozek R. A., 2007, Phys. Rev. D, 75, 103508
Lee Y.-W., Chung C., Kang Y., Jee M. J., 2020, ApJ, 903, 22
Leistedt B., Mortlock D. J., Peiris H. V., 2016, MNRAS, 460, 4258
Linder E. V., Mitra A., 2019, Phys. Rev. D, 100, 043542
LSST Science Collaboration, 2009, preprint (arXiv:0912.0201)
Lueckmann J.-M., Boelts J., Greenberg D., Goncalves P., Macke J., 2021, Proc. 24th Int. Conf. Artif. Intell. Stat., Benchmarking Simulation-Based Inference. PMLR, Cambridge, MA, p. 343
Ma C., Corasaniti P.-S., Bassett B. A., 2016, MNRAS, 463, 1651
Malmquist K. G., 1922, Medd. Fran Lunds Astron. Obs. Ser. I, 100, 1
Malmquist K. G., 1925, Medd. Fran Lunds Astron. Obs. Ser. I, 106, 1
Mandel K. S., Wood-Vasey W. M., Friedman A. S., Kirshner R. P., 2009, ApJ, 704, 629
Mandel K. S., Narayan G., Kirshner R. P., 2011, ApJ, 731, 120
Mandel K. S., Scolnic D. M., Shariff H., Foley R. J., Kirshner R. P., 2017, ApJ, 842, 93
Mandel K. S., Thorp S., Narayan G., Friedman A. S., Avelino A., 2022, MNRAS, 510, 3939
March M. C., Trotta R., Berkes P., Starkman G. D., Vaudrevange P. M., 2011, MNRAS, 418, 2308
Masserano L., Dorigo T., Izbicki R., Kuusela M., Lee A. B., 2022, preprint (arXiv:2205.15680)
Miller B. K., Cole A., Louppe G., Weniger C., 2020, preprint (arXiv:2011.13951)
Miller B. K., Cole A., Forré P., Louppe G., Weniger C., 2022, J. Open Source Softw., 7, 4205

---

[18] https://github.com/kosiokarchev/clipppy
[19] https://github.com/kosiokarchev/phytorch
[20] https://www.pytorchlightning.ai/

Neyman J., Jeffreys H., 1937, Philos. Trans. R. Soc. Lond. Ser. Math. Phys. Sci., 236, 333
Nicolas N. et al., 2021, A&A, 649, A74
Paszke A. et al., 2019, in Wallach H., Larochelle H., Beygelzimer A., d'Alché-Buc F., Fox E., Garnett R., eds, Advances in Neural Information Processing Systems 32. Curran Associates, Inc., Canada, p. 8024
Peebles P. J. E., 1993, Principles of Physical Cosmology. Princeton Univ. Press, Princeton, NJ
Perlmutter S. et al., 1997, ApJ, 483, 565
Perlmutter S. et al., 1999, ApJ, 517, 565
Phillips M. M., 1993, ApJ, 413, L105
Phillips M. M. et al., 2019, PASP, 131, 014001
Planck Collaboration I, 2020, A&A, 641, A1
Popovic B., Brout D., Kessler R., Scolnic D., 2021a, preprint (arXiv:2112.04456)
Popovic B., Brout D., Kessler R., Scolnic D., Lu L., 2021b, ApJ, 913, 49
Pskovskii Y. P., 1967, Sov. Astron., 11, 63
Pskovskii I. P., 1977, Sov. Astron., 21, 675
Pskovskii Y. P., 1984, Sov. Astron., 28, 658
Rahman W., Trotta R., Boruah S. S., Hudson M. J., van Dyk D. A., 2022, MNRAS, 514, 139
Revsbech E. A., Trotta R., van Dyk D. A., 2018, MNRAS, 473, 3969
Riess A. G. et al., 1998, AJ, 116, 1009
Roberts E., Lochner M., Fonseca J., Bassett B. A., Lablanche P.-Y., Agarwal S., 2017, J. Cosmol. Astropart. Phys., 2017, 036
Rubin D. et al., 2015, ApJ, 813, 137
Sako M. et al., 2018, PASP, 130, 064002
Saunders C. et al., 2018, ApJ, 869, 167
Schlafly E. F., Finkbeiner D. P., 2011, ApJ, 737, 103
Scolnic D. M. et al., 2018, ApJ, 859, 101
Scolnic D. et al., 2022, ApJ, 938, 113
Shariff H., Dhawan S., Jiao X., Leibundgut B., Trotta R., van Dyk D. A., 2016a, MNRAS, 463, 4311
Shariff H., Jiao X., Trotta R., van Dyk D. A., 2016b, ApJ, 827, 1
Silverman J. M. et al., 2012, MNRAS, 425, 1789
Sisson S., Fan Y., Beaumont M., 2018, Handbook of Approximate Bayesian Computation, 1st edn. Chapman and Hall/CRC
Talts S., Betancourt M., Simpson D., Vehtari A., Gelman A., 2020, preprint (arXiv:1804.06788)
Taylor G., Lidman C., Tucker B. E., Brout D., Hinton S. R., Kessler R., 2021, MNRAS, 504, 4111
Tripp R., 1997, A&A, 325, 871
Tripp R., 1998, A&A, 331, 815
Weinberg S., 2008, Cosmology, illustrated edition. Oxford Univ. Press, Oxford
Weyant A., Schafer C., Wood-Vasey W. M., 2013, ApJ, 764, 116
Wojtak R., Davis T. M., Wiis J., 2015, J. Cosmol. Astropart. Phys., 2015, 025

## APPENDIX A: P–P PLOTS

A P–P plot is a diagnostic tool which compares credibilities $\gamma(\mathbf{\Theta}_0, \mathbf{d})$, as defined in equation (15), with their cumulative distribution, $F(\gamma) \equiv \int_0^\gamma p(\gamma') \, d\gamma'$, where $p(\gamma')$ is implied by a distribution of $\mathbf{\Theta}_0$ and $\mathbf{d}$. The interpretation of $F(\gamma)$ is the frequency with which parameters are covered by credible regions and is hence called the 'empirical coverage'. In practice, it is constructed by drawing a sample for $\mathbf{\Theta}_0$ and $\mathbf{d}$, evaluating the approximate posterior across the whole parameter space, determining $\Gamma_\Theta(\mathbf{\Theta}_0, \mathbf{d})$ (from equation 16), and integrating over it, to determine the approximate credibility associated with $\mathbf{\Theta}_0$, as illustrated in the top panel of Fig. 3. This results in a sample from $p(\gamma)$.

### A1 Bayesian coverage

In the LFI literature, notably Hermans et al. (2022), $\mathbf{\Theta}_0$, $\mathbf{d}$ are taken to follow the joint distribution $p(\mathbf{\Theta}_0, \mathbf{d})$ defined by a forward model

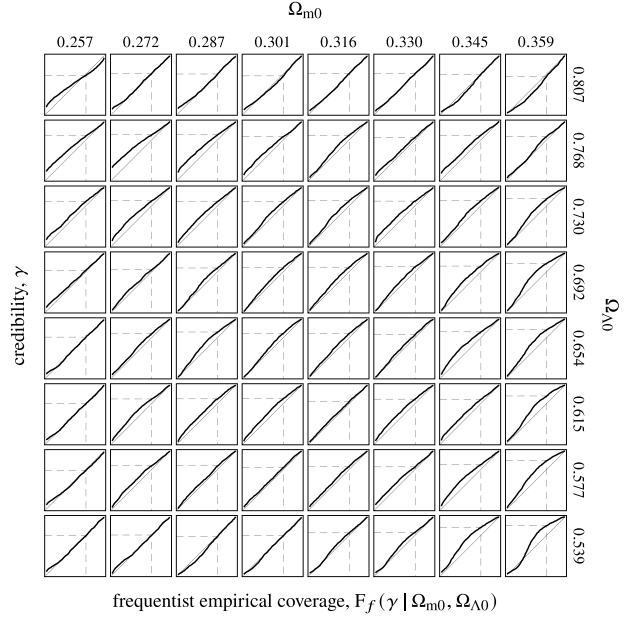

**Figure A1.** Frequentist P–P plots; i.e. at fixed cosmology (cf. the middle panel of Fig. 3), created with the final cosmology-only posterior approximation for $10^5$ SNe Ia. Each panel corresponds to a pixel in the map of Fig. 8b with cosmological parameter values indicated at the top and right edges (only some pixels are shown). The colour values in Fig. 8b correspond to the credibilities indicated with a horizontal dashed line: those that achieve an empirical coverage of 68.4 per cent (vertical dashed line).

(simulator). In this set-up, the probability density of $\gamma$ is

$$p_B(\tilde{\gamma}) = \iint \delta(\tilde{\gamma} - \gamma(\mathbf{\Theta}_0, \mathbf{d})) \, p(\mathbf{\Theta}_0, \mathbf{d}) \, d\mathbf{\Theta}_0 \, d\mathbf{d}, \tag{A1}$$

and so its cumulative distribution is, after switching the order of integration,

$$F_B(\tilde{\gamma}) = \iint \left[ \int_0^{\tilde{\gamma}} \delta(\gamma' - \gamma(\mathbf{\Theta}_0, \mathbf{d})) \, d\gamma' \right] p(\mathbf{\Theta}_0, \mathbf{d}) \, d\mathbf{\Theta}_0 \, d\mathbf{d}$$

$$= \iint_{\{\mathbf{\Theta}_0, \mathbf{d}: \, \tilde{\gamma} > \gamma(\mathbf{\Theta}_0, \mathbf{d})\}} p(\mathbf{\Theta}_0, \mathbf{d}) \, d\mathbf{\Theta}_0 \, d\mathbf{d} \tag{A2}$$

where the step resulting from integrating the delta function has been written as an appropriate region of integration. This expression corresponds to the probability mass, averaged over all possible data, *of the true posterior* in the regions that have credibility $\tilde{\gamma}$ under the approximate posterior:

$$F_B(\tilde{\gamma}) = \mathbb{E}_{p(\mathbf{d})} \left[ \int_{\Gamma_{\Theta_0}(\gamma^{-1}(\tilde{\gamma}, \mathbf{d}), \mathbf{d})} p(\mathbf{\Theta}_0 \mid \mathbf{d}) \, d\mathbf{\Theta}_0 \right], \tag{A3}$$

with $\gamma^{-1}(\tilde{\gamma}, \mathbf{d})$ an inversion of $\gamma(\mathbf{\Theta}_0, \mathbf{d})$ for a fixed $\mathbf{d}$.

Thus, we see that when $q(\mathbf{\Theta} \mid \mathbf{d}) \to p(\mathbf{\Theta} \mid \mathbf{d})$, the integral will be constant across all data sets and equal to the credibility, $\tilde{\gamma}$, and so $F_B(\tilde{\gamma}) \to \tilde{\gamma}$. In other words, if the estimator $q$ is well calibrated, the empirical coverage will match the credibility, and so the P–P plot will be a diagonal line.[21] On the other hand, when $q$ is 'wider' than the true posterior, i.e. the ratio estimator is conservative, the area of integration will be larger, and so $F_B(\tilde{\gamma}) > \tilde{\gamma}$. An example of a

---

[21] Note that this arises only in the so-called Bayesian optimal setting, in which the posterior is derived using the model used to generate the analysed data, and it is crucial to average the credibilities over the exact joint model.

Bayesian P–P plot is Fig. 7, where the empirical coverage has been averaged over the entire training domain of the inference network that is being validated.

However, a diagonal Bayesian P–P plot is not a sufficient condition for establishing convergence of $q$ on to the true posterior because of the data-averaging in equation (A3): one can imagine $q$ conspiring to be conservative for some *data* and overconfident for others in ways that exactly cancel.

## A2 Frequentist coverage testing and calibration

In frequentist statistics, the parameters are fixed,[22] and the data follows the sampling distribution, $p(\boldsymbol{d} \mid \boldsymbol{\Theta})$. A frequentist P–P plot, therefore, depicts

$$F_{\mathrm{f}}(\tilde{\gamma} \mid \boldsymbol{\Theta}_0) = \int \left[ \int_0^{\tilde{\gamma}} \delta(\gamma' - \gamma(\boldsymbol{\Theta}_0, \boldsymbol{d})) \, \mathrm{d}\gamma' \right] p(\boldsymbol{d} \mid \boldsymbol{\Theta}_0) \, \mathrm{d}\boldsymbol{d}. \quad (A4)$$

Unlike $F_{\mathrm{B}}(\tilde{\gamma})$, the frequentist empirical coverage $F_{\mathrm{f}}(\tilde{\gamma} \mid \boldsymbol{\Theta}_0)$ needs to be determined by sampling even for the exact posterior due to the effect of the prior, which is integrated out in equation (A2). Furthermore, while the credible regions of the true posterior are guaranteed to have perfect coverage *when averaged over the joint inference model*, this is not necessarily true for fixed $\boldsymbol{\Theta}_0$ if the prior in $\boldsymbol{\Theta}$ is not uniform. Still, we can use the approximate Bayesian credibilities to construct confidence regions with exact coverage.

To this end, we define the 'required credibility' $\hat{\gamma}(\boldsymbol{\Theta}_0, \tilde{\gamma})$ as the credibility of approximate credible regions which have frequentist coverage with frequency $\tilde{\gamma}$, i.e. as the solution of

$$F_{\mathrm{f}}(\hat{\gamma} \mid \boldsymbol{\Theta}_0) = \int_0^{\hat{\gamma}} p(\gamma' \mid \boldsymbol{\Theta}_0) \, \mathrm{d}\gamma' = \tilde{\gamma}. \quad (A5)$$

Thus, $\hat{\gamma}$ is simply the $\tilde{\gamma}^{\mathrm{th}}$ quantile in the frequentist P–P plot: see the middle panel of Fig. 3 and the concrete results for cosmological parameter inference with TMNRE in Fig. A1.

Now we can trivially construct frequentist confidence regions with exact coverage using $\hat{\gamma}$. For a given data set, we evaluate the approximate posterior and from it $\gamma(\boldsymbol{\Theta}_0, \boldsymbol{d})$, which we now treat as a test statistic. If the credibility associated with a given parameter value $\boldsymbol{\Theta}_0$ is lower than the required credibility, the parameter is included in the confidence region:

$$C_{\boldsymbol{\Theta}_0}(\boldsymbol{d}, \tilde{\gamma}) \equiv \{\boldsymbol{\Theta}_0 : \gamma(\boldsymbol{\Theta}_0, \boldsymbol{d}) \leq \hat{\gamma}(\boldsymbol{\Theta}_0, \tilde{\gamma})\}. \quad (A6)$$

Thus, if the approximate posterior already has exact frequentist coverage ($\hat{\gamma}(\boldsymbol{\Theta}_0, \tilde{\gamma}) = \tilde{\gamma}$), no modification is necessary, but otherwise, undercovered parameters ($\hat{\gamma}(\boldsymbol{\Theta}_0, \tilde{\gamma}) > \tilde{\gamma}$) are included more often. A demonstration is shown in the bottom panel of Fig. 3.

## APPENDIX B: LINEARIZED MARGINAL BAHAMAS

The BAHAMAS model, as described in Table 1 and depicted in Fig. 1, is non-Gaussian only in the priors of some global parameters. Furthermore, the only non-linearity is present in the distance modulus, which has a complicated dependence on the redshift. If the redshift of each supernova is fixed (e.g. with spectroscopic observations), or the uncertainties $\sigma_z^s \equiv (1 + \hat{z}^s)\sigma_z$ are small enough so that they can be propagated at linear order, the latent model becomes Gaussian as well, so that it can be analytically marginalized; i.e. the SN Ia plate can be collapsed to a single normal distribution giving the prior of the latent parameters, conditioned on the global parameters, namely

$$\begin{bmatrix} m^s \\ x_1^s \\ c^s \end{bmatrix} \sim \mathcal{N}\left( \bar{\boldsymbol{d}}_{\mathrm{p}}^s \equiv \begin{bmatrix} \bar{m}^s \\ \bar{x}_1 \\ \bar{c} \end{bmatrix}, \, \boldsymbol{\Sigma}_{\mathrm{p}}^s \equiv \begin{bmatrix} (R_{\mathrm{m}}^s)^2 & -\alpha R_{x_1}^2 & \beta R_{\mathrm{c}}^2 \\ -\alpha R_{x_1}^2 & R_{x_1}^2 & 0 \\ \beta R_{\mathrm{c}}^2 & 0 & R_{\mathrm{c}}^2 \end{bmatrix} \right) \quad (B1)$$

with

$$\bar{m}^s = \bar{M}_0 - \alpha \bar{x}_1 + \beta \bar{c} + \mu\left(\mathcal{C}, \hat{z}^s\right), \quad (B2)$$

$$\left(R_{\mathrm{m}}^s\right)^2 = \sigma_{\mathrm{res}}^2 + (\alpha R_{x_1})^2 + (\beta R_{\mathrm{c}})^2 + \left(\left.\frac{\partial \mu}{\partial z}\right|_{\mathcal{C}, \hat{z}^s} \sigma_z^s \right)^2. \quad (B3)$$

Note that even though $\partial\mu/\partial z$ is evaluated at a fixed set of redshifts, it has to be recalculated at each new draw of the cosmological parameters.

The prior covariance from equation (B1) is added to the data covariance, $\hat{\boldsymbol{\Sigma}}$, to obtain the marginal covariance of the data, given the global parameters:

$$\boldsymbol{d} \mid \boldsymbol{\Theta} \sim \mathcal{N}(\bar{\boldsymbol{d}}_{\mathrm{p}}, \boldsymbol{\Sigma}_{\mathrm{p}} + \hat{\boldsymbol{\Sigma}}), \quad (B4)$$

where $\boldsymbol{\Theta} \equiv \{\mathcal{C}, \sigma_z, \alpha, \beta, \bar{M}_0, \sigma_{\mathrm{res}}, \bar{x}_1, R_{x_1}, \bar{c}, R_{\mathrm{c}}\}$ are the global parameters of the model, which the marginal prior mean, $\bar{\boldsymbol{d}}_{\mathrm{p}} \equiv \mathrm{concat}_s\left(\bar{\boldsymbol{d}}_{\mathrm{p}}^s\right)$, and covariance, $\boldsymbol{\Sigma}_{\mathrm{p}} \equiv \mathrm{diag}_s\left(\boldsymbol{\Sigma}_{\mathrm{p}}^s\right)$, are functions of. As March et al. (2011, equation C19) show, evaluating this likelihood quickly, i.e. without inverting the large matrix $\boldsymbol{\Sigma}_{\mathrm{p}} + \hat{\boldsymbol{\Sigma}}$, is possible only up a normalization factor that depends on $\boldsymbol{\Theta}$ and therefore needs to be recomputed at every MCMC step, incurring a determinant calculation of $\mathcal{O}((3N)^3)$ complexity.[23] In the simplified case of independent likelihoods for each SN Ia, i.e. a block-diagonal $\hat{\boldsymbol{\Sigma}}$ (in addition to $\boldsymbol{\Sigma}_{\mathrm{p}}$), which we consider, however, the complexity is anyway reduced to $\mathcal{O}(3^3 \times N)$.

Thus, in its linearized marginal version and assuming independent likelihoods for each SN Ia, BAHAMAS can easily be analysed with MCMC in $\mathcal{O}(10)$ dimensions by inverting a block-diagonal matrix, as was done for the comparison in Fig. 9.

### B1 The latent parameters a posteriori

Even though the latent parameters do not appear explicitly in marginal BAHAMAS, one can derive posteriors for them, considering, initially, the conditional posterior

$$\boldsymbol{m}, \boldsymbol{x}_1, \boldsymbol{c} \mid \boldsymbol{\Theta}, \boldsymbol{d} \sim \mathcal{N}\Bigg( \hat{\boldsymbol{\mu}}_{\bar{d}} = \left(\hat{\boldsymbol{\Sigma}}^{-1} + \boldsymbol{\Sigma}_{\mathrm{p}}^{-1}\right)^{-1} \left(\boldsymbol{\Sigma}_{\mathrm{p}}^{-1} \bar{\boldsymbol{d}}_{\mathrm{p}} + \hat{\boldsymbol{\Sigma}}^{-1} \boldsymbol{d}\right),$$
$$\hat{\boldsymbol{V}}_{\bar{d}} = \left(\hat{\boldsymbol{\Sigma}}^{-1} + \boldsymbol{\Sigma}_{\mathrm{p}}^{-1}\right)^{-1} \Bigg), \quad (B5)$$

which is the posterior of an a priori Gaussian variable under Gaussian likelihood. The marginal posterior for the latent variables is then the marginalization of this expression over the marginal posterior of the global parameters, which can be obtained in a post-processing step of the global MCMC run, in which the means and variances from equation (B5), evaluated for each sampled $\boldsymbol{\Theta}$, are averaged. This procedure was used to calculate the values on the abscissa of Fig. 11.

---

[22] In this context, fixed are only the parameters of interest, while all other nuisance and latent variables are still marginalized.

[23] A stochastic estimate based on an augmented conjugate gradient technique can be obtained in linear time, as detailed by Gardner et al. (2021).

This paper has been typeset from a T<sub>E</sub>X/L<sup>A</sup>T<sub>E</sub>X file prepared by the author.